\documentclass[11pt]{article}
\usepackage{hyperref}
\usepackage{verbatim}
\usepackage{fullpage}
\usepackage{amsmath}
\usepackage{latexsym}
\usepackage{revsymb}
\usepackage{yfonts}

\usepackage{natbib}
\usepackage{amsfonts}
\usepackage{amsmath}
\usepackage{amssymb}
\usepackage{amsthm}
\usepackage{graphicx}
\usepackage{bm}
\usepackage{bbm}

\newcommand{\be}{\begin{eqnarray} \begin{aligned}}
\newcommand{\ee}{\end{aligned} \end{eqnarray} }
\newcommand{\benn}{\begin{eqnarray*} \begin{aligned}}
\newcommand{\eenn}{\end{aligned} \end{eqnarray*} }

\newcommand{\bc}{\begin{center}}
\newcommand{\ec}{\end{center}}

\newcommand{\ran}{\rangle}
\newcommand{\lan}{\langle}
\newcommand{\id}{\mathbb{I}}

\newcommand{\Tr}{\mathop{\mathrm{Tr}}\nolimits}

\newcommand{\pr}{\prime}

\newcommand{\iu}{i}

\newcommand{\e}{\mathrm{e}}

\newtheorem{theorem}{Theorem}[section]

\newtheorem{lemma}[theorem]{Lemma}
\newtheorem{definition}[theorem]{Definition}

\newtheorem{corollary}[theorem]{Corollary}

\newcommand{\hil}{\mathcal{H}}

\newcommand{\spann}{\mathop{\mathrm{span}}\nolimits}  

\newcommand{\dfdas}{\stackrel{\textrm{\tiny def}}{=}}

\newcommand{\ds}{\displaystyle}

\usepackage{amsfonts}

\def\Complex{\mathbb{C}}
\def\Natural{\mathbb{N}}
\def\id{\mathbb{I}}

\def\01{\{0,1\}}

\newcommand{\ket}[1]{|#1\rangle}
\newcommand{\bra}[1]{\langle#1|}
\newcommand{\outp}[2]{|#1\rangle\langle#2|}
\newcommand{\ketbra}[2]{|#1\rangle\langle#2|}

\newcommand{\inp}[2]{\langle{#1}|{#2}\rangle} 

\newcommand{\rank}{\operatorname{rank}}

\newcommand{\mX}{\mathcal{X}}
\newcommand{\mY}{\mathcal{Y}}
\newcommand{\mB}{\mathcal{B}}

\newenvironment{sdp}[2]{
\smallskip
\begin{center}
\begin{tabular}{ll}
#1 & #2\\
subject to
}
{
\end{tabular}
\end{center}
\smallskip
}

\newcommand{\ens}{\mathcal{E}}
\newcommand{\Y}{|\mY|}

\begin{document}

\title{{\sf State Discrimination with Post-Measurement Information}}

\author{Manuel A. Ballester$^1$, Stephanie Wehner$^1$ and Andreas Winter$^2$\\
\textit{$^1$CWI, Kruislaan 413, 1098 SJ Amsterdam, The Netherlands}\\
\textit{$^2$Department of Mathematics, University of Bristol, Bristol BS8 1TW, U.~K.}}

\date{1 August 2006}

\maketitle

\begin{abstract}
 We introduce a new state discrimination problem in which we are given additional 
 information about the state after the measurement, or more generally, after a quantum
 memory bound applies. 
 In particular, the following special case plays an important role in quantum cryptographic
 protocols in the bounded storage model:
 Given a string $x$ encoded in an unknown basis chosen from a set of mutually unbiased bases, 
 you may perform any measurement, but then store at most $q$ qubits of quantum information.
 Later on, you learn which basis was used. How well can you compute a function
 $f(x)$ of $x$, given the initial 
 measurement outcome, the $q$ qubits and the additional basis information? 
 We first show a lower bound on the success probability 
 for any balanced function, and any number of mutually unbiased 
 bases, beating the naive strategy of simply guessing the basis. 
 We then show that for two bases, any Boolean function $f(x)$ can be computed perfectly if
 you are allowed to store just a single qubit, independent of the number of
 possible input strings $x$. 
 However, we show how to construct \emph{three} bases, such that you need to store \emph{all}
 qubits in order to compute $f(x)$ perfectly.
 We then investigate how much advantage the additional basis information 
 can give for a Boolean
 function. To this end, we prove optimal bounds for the success 
 probability for the AND 
 and the XOR function for up to three mutually unbiased bases. 
 Our result shows that the gap in success probability can be maximal: 
 \emph{without} the basis information, you can never do better than guessing 
 the basis, but \emph{with} this information,
 you can compute $f(x)$ perfectly. We also exhibit an example 
 where the extra information
 does not give any advantage at all. 
\end{abstract}

\thispagestyle{empty}

\renewcommand{\contentsname}{}
\setcounter{tocdepth}{1}
\tableofcontents

\section{Introduction}\label{introduction}

State discrimination with post-measurement information concerns the following
task: Consider an ensemble of quantum states,
${\cal E} = \{ p_{yb},\rho_{yb} \}$, with double indices
$yb \in {\cal Y}\times{\cal B}$, and a number $q\geq 0$.
Suppose 
Alice sends Bob the state $\rho_{yb}$, where she alone knows indices $y$ and $b$.
Bob can perform any 
measurement on his system, but then store at most $q$ qubits (i.e.~a Hilbert
space of dimension $2^q$). Afterwards, Alice tells him $b$. Bob's goal is now
to approximate $y$ as accurately as possible. Here, this means that
he has to make a guess $\hat{Y}$, maximizing the success probability
\[
  p_{\rm succ} = \sum_{yb} p_{yb} \Pr\{ \hat{Y} = y | \text{state } \rho_{yb} \}.
\]

For $|{\cal B}|=1$, i.e.~no available post-measurement information, $q$ is irrelevant
and Bob's task is to discriminate among states $\rho_y$, a problem studied
since the early days of quantum information science~\cite{helstrom}.
On the other hand, if the $\rho_{yb}$ all commute, the fact that $b$ comes
later -- and also the magnitude of $q$ -- plays no role as Bob
can always measure in the common eigenbasis of the states without losing
any information.

Hence, a particular case of the general problem that isolates the aspect
of the timing between measurements and side-information is one where for
each fixed $b$, the states $\rho_{yb}$ are mutually orthogonal:
\begin{equation}
  \label{rhoyb-orth}
  \forall b \forall y\neq z\quad \rho_{yb} \perp \rho_{zb}.
\end{equation}
Then the difficulty for Bob and the nontrivial dependence of his probability
of success on $q$ derive from the possibility of non-commuting eigenbases
of the sets $\{\rho_{yb} \}$ for different $b$. While for given $b$
he can distinguish perfectly between the $\rho_{yb}$, the quantum mechanical
measurement-disturbance principle reduces the success probability if
this side-information is delayed.

In this paper, we focus for the most part on a
special case that is of central importance to
existing protocols in the bounded quantum storage model~\cite{serge:bounded}.
The security of such protocols rests on the realistic assumption that a dishonest
player cannot store more than $q$ qubits for long periods of time. In this model,
even bit commitment and oblivious transfer can be implemented securely, which
is otherwise known to be
impossible~\cite{lo&chau:bitcom,mayers:trouble,kretschmann&werner:impossible}.
In particular, we are interested in the following question:
Consider a function $f:{\cal X}\rightarrow{\cal Y}$ between 
finite sets, and a set of mutually unbiased bases ${\cal B}$, given by unitaries 
$U_0=\id,U_1,\ldots,U_{|{\cal B}|-1}$ on a Hilbert space
with basis $\{\ket{x}:x\in{\cal X}\}$. Alice chooses a string $x$ and a basis $b$
where $xb$ is drawn from the distribution $P_{X,B}$, prepares the state $U_b\ket{x}$ 
and sends it to Bob. 
When Bob receives the state, he may perform any measurement. Afterwards, however,
he can store at most $q$ qubits of quantum information. Later, Alice announces which basis
she had chosen. Bob's task is now to predict $y=f(x)$ as accurately as possible.
That means that the states in our problem 
are now given by
\[
  \rho_{yb} = \sum_{x\in f^{-1}(y)} P_{X|B}(x) U_b \outp{x}{x} U_b^\dagger.
\]
The only difference to Eq.~(\ref{rhoyb-orth}) is that now we demand the
mutual unbiasedness of the joint eigenbases of the $\{ \rho_{yb} \}$ for
different $b$.
How well can Bob compute $f(x)$ given the classical outcome of his earlier
measurement, the $q$ qubits and the additional basis information? 
In the context of cryptographic protocols~\cite{serge:bounded},
Bob is a dishonest player who tries to learn some
function of the encoded string conditioned on the fact that he will later
learn the basis and the function.
In the oblivious transfer protocol of~\cite{serge:bounded}, Alice uses two 
mutually unbiased bases, and secretly chooses a 
function from a set of predetermined functions. She then tells
Bob which function he should evaluate together
with the basis information $b$. This makes the protocols more
complicated and so one might
wonder whether it is possible to use a fixed Boolean function
instead, a question which stood at the beginning of the current
investigation (see~\cite{serge:bounded}, section 3.6 in the {\tt arXiv}
version, where the XOR of an even number of bits is shown to be insufficient).
However, we show that this is
not possible in the suggested protocol. In particular, we show that for two bases 
and \emph{any} Boolean function $f$, Bob can succeed with probability
at least $1/2 + 1/(2\sqrt{2})$, even if
he cannot store any qubits at all. Surprisingly, it also turns out that 
Bob can determine $f(x)$ perfectly, if he can store 
just a \emph{single} qubit. We show that 
one qubit is sufficient no matter how long the input string 
$x$ actually is. Behind our proof, there is an algebraic framework
that allows us, in principle, to determine the minimal quantum
memory resources required and the optimal strategy to
succeed with probability $1$ for any number of bases and any function $f$.
However, it turns out that we can construct \emph{three} bases, such
that Bob needs to store \emph{all} qubits in order to compute a boolean 
function perfectly.

In general, we also show a lower bound on Bob's 
optimal success probability for any balanced function 
$f: \mX \rightarrow \mY$, and any number of mutually unbiased
bases if he cannot store any qubits. 
Our bound is strictly better than what Bob could achieve by guessing the basis. 

Our problem also has an interpretation in the light of communication complexity. Suppose
Alice is given $b$, and Bob is given the state $\rho_{yb}$. 
If classical communication is free, what is the minimal number of qubits Bob needs to
communicate to Alice such that Alice learns $y$? It turns out that if there exists a 
strategy for Bob to compute $y$ in our original task while storing only $q$ qubits,
he will also need to send exactly $q$ qubits, and his classical measurement outcome, to allow 
Alice to learn $y$: Alice now simply 
performs the measurement Bob would have done in our original task after he received $b$.

It is an interesting problem to consider how much the extra basis information
helps Bob to compute $f(x)$. 
To this end, we first examine how well Bob can compute the AND
and XOR of $x$ \emph{without} using the additional basis
information. We prove optimal bounds
for computing the AND and XOR function on a string of length
$n$ for two and three mutually unbiased 
bases. In particular, we show that for two mutually unbiased bases and the 
XOR function on strings of even length, Bob's probability of 
success is at most $3/4$, and there exists a strategy
which achieves it.  This means that his trivial strategy of
guessing the basis and taking the measurement outcome 
in that basis to be the real answer, is optimal.
Interestingly, adding the third basis does not change
his success probability of $3/4$, whereas intuitively one would expect it to be lower. 
Surprisingly, for three bases, if we choose a non-uniform prior distribution over the 
strings of length $n$, it actually becomes harder for Bob
to compute the XOR. We show that there exists
a non-uniform distribution such that he can never succeed more than 
using the trivial strategy of guessing the outcome.
No measurement he can perform will give him any more information.
We then examine the case that the length of the string $n$ is
odd. Here, Bob can succeed only with probability $1/2 + 1/(2\sqrt{2})$ which is optimal.
We prove that for \emph{any} Boolean function $f$,
Bob's probability of success is upper bounded by 
$1/2 + 1/(2\sqrt{|\mB|})$ if he does not receive any basis information. 

We then examine how well Bob can do \emph{with} the additional basis information. 
We show that for the XOR function on strings of even
length, Bob can now compute the value of the function perfectly 
and give an explicit measurement strategy for Bob.
For two bases, this means that the gap can be maximal:
\emph{without} the basis information Bob cannot do better
than the trivial strategy of guessing the basis, however,
\emph{with} this extra information Bob always succeeds.
It also means that the gap can be minimal: For the XOR on strings
of odd length the extra information does not help Bob at all.
For three bases we obtain the maximum gap only for a
non-uniform prior. Finally, we also give an optimal strategy
for computing the AND from the given state and the post-measurement information.

\subsection{Related work}

State discrimination itself has received considerable
attention in the past: Alice prepares a quantum 
state drawn from a collection of possible quantum states.
Bob's goal is now to determine the identity of the state.
The new twist in the present work is that after the measurement,
or more generally after a memory bound applies, he is given
additional information.
For the case of only two (mixed) states, the optimal
measurement for traditional state discrimination was found by
Helstrom~\cite{helstrom}. The case of multiple (mixed) states was
already considered by Holevo~\cite{holevo:maxState} and  
Yuen, Kennedy and Lax~\cite{yuen:maxState} in the 70's, and  
they have given the necessary conditions for a measurement to be optimal. Yuen 
et al. also showed these conditions to be sufficient and
demonstrated that the problem of finding the 
optimal measurement can be expressed as convex optimization problem.
Discriminating between multiple mixed states remains a 
difficult problem and it is usually hard to derive
explicit measurements and bounds from these conditions.
Optimal measurements are known only for 
special state sets, which satisfy certain symmetry
properties~\cite{eldar:symmetric,ban:symmetric,barnett:symmetric}. 

Many convex optimization problems can be solved using semidefinite programming.
Eldar~\cite{eldar:sdp} and 
Eldar, Megretski and Verghese~\cite{eldar:sdpDetector}
used semidefinite programming to solve state discrimination
problems, which is one of the
techniques we will also use here. The square-root
measurement~\cite{hausladen:pgm} (also called pretty good measurement)
is an easily constructed measurement to distinguish quantum states,
however, it is only optimal for 
very specific sets of states~\cite{eldar:pgm,eldar:symmetric}.
Mochon constructed specific pure state discrimination problems for which the square-root
measurement is optimal~\cite{mochon:pgm}.
We will use a variant of the square-root measurement as well.
Furthermore, our problem is related to the task of state
filtering~\cite{bergou:filtering0,bergou:filtering, bergou:filtering2}
and state classification~\cite{wang:classification}.
Here, Bob's goal is to determine whether a given state is 
either a specific state or one of several other possible 
states, or, more generally, which subset of states a 
given state belongs to. Our scenario differs, because
we deal with mixed states and Bob 
is allowed to use post-measurement information. Much more is known about pure state 
discrimination problems and the case
of unambiguous state discrimination where we are not allowed to make an error.
Since we concentrate on mixed states, we refer
to~\cite{bergou:survey} for an excellent survey on 
the extended field of state discrimination.

Regarding state discrimination with post-measurement information,
special instances of the general problem have occurred in the literature
under the heading ``mean king's problem''~\cite{aharonov&englert:meanking,%
klappenecker&roetteler:meanking},
where the stress was on the usefulness of entanglement. Furthermore,
it should be noted that prepare-and-measure quantum key distribution
schemes of the BB84 type also lead to special cases of this problem:
When considering optimal individual attacks, the eavesdropper is faced
with the task of extracting maximal information about the raw key bits, encoded
in an unknown basis, that she learns later during basis reconciliation.

\section{Preliminaries}\label{prelim}

\subsection{Notation and tools}

We will need the following notions. The Bell basis is given by the vectors
$\ket{\Phi^{\pm}} = (\ket{00} \pm \ket{11})/\sqrt{2}$ and $\ket{\Psi^{\pm}}
                  = (\ket{01} \pm \ket{10})/\sqrt{2}$.
Furthermore, let $f^{-1}(y) = \{x \in \mX|f(x) = y\}$.
We say that a function $f$ is balanced if and only if any element
in the image of $f$ is generated by equally many elements in the pre-image of $f$, i.e.
there exists a $k \in \Natural$ such that $\forall y \in \mY: |f^{-1}(y)| = k$.
We also use the notation 
$[m] = \{1,\ldots,m\}$. $A^{\dagger}$ is the conjugate transpose of matrix $A$.
A \emph{positive semidefinite} $n \times n$
matrix $A$ is a Hermitian matrix such that 
$x^*Ax \geq 0$ for all $x \in \Complex^n$~\cite{horn&johnson:ma}.
$A$ is said to be \emph{positive definite}, if it is 
positive semidefinite and  $x^*Ax = 0$ implies $x=0$.
We use $A \geq 0$ and $A > 0$ to indicate that $A$ is 
positive semidefinite and positive definite, respectively.
Finally, $\|A\|_1 = \Tr\sqrt{A^\dagger A}$ is the trace norm.
The first tool we use is the following well-known result.

\begin{theorem}[Helstrom~\cite{helstrom}]\label{helstrom}
  Suppose we are given states $\rho_0$ with probability $q$, and $\rho_1$
  with probability $1-q$. Then the probability to determine whether the state was
  $\rho_0$ and $\rho_1$ is at most
  \[
    p = \frac{1}{2}\left[1 + \|q\rho_0 - (1-q) \rho_1\|_1\right].
  \]
  The measurement that achieves $p$ is given by $M_0$, 
  and $M_1 = \id - M_0$, where $M_0$ is the projector onto the positive eigenspace of 
  $q\rho_0 - (1-q)\rho_1$.
  \qed
\end{theorem}

Secondly, we will make use of semidefinite programming, which is a special
case of convex optimization. We refer to~\cite{boyd:book} for an in-depth introduction.
The goal of semidefinite programming is to solve he 
following semidefinite program (SDP) in terms of the variable $X \in S^n$
\begin{sdp}{maximize}{$\Tr(CX)$}
&$\Tr(A_iX) = b_i, i = 1,\ldots,p$,
and $X \geq 0$
\end{sdp}
for given matrices $C,A_1,\ldots,A_p \in S^n$ where 
$S^n$ is the space of symmetric $n \times n$ matrices.
$X$ is called \emph{feasible}, if it satisfies all constraints.
An important aspect of semidefinite programming is duality. 
Intuitively, the idea behind Lagrangian duality is to extend the objective function
(here $\Tr(CX)$) with a weighted sum of the constraints 
in such a way, that we will be penalized
if the constraints are not fulfilled. The weights then 
correspond to the dual variables.
Optimizing over these weights then gives rise to the \emph{dual problem}.
The original problem is called the \emph{primal problem}.
Let $d'$ denote the optimal value of the dual problem, and $p'$ the optimal value
of the primal problem stated above. Weak duality says that $d' \geq p'$.
In particular, if we have $d' = p'$ for a feasible dual and primal solution
respectively, we can conclude that both solutions are optimal.

We will also need the notion of mutually unbiased bases,
which was introduced in~\cite{wootters:mub}. 
The following definition closely follows the one given in~\cite{boykin:mub}.
\begin{definition} \label{def-mub}
Let $B_1 = \{\ket{\phi_1},\ldots,\ket{\phi_d}\}$ and $B_2 =
\{\ket{\psi_1},\ldots,\ket{\psi_d}\}$ be two orthonormal bases in
a $d$ dimensional Hilbert space. They are said to be
\emph{mutually unbiased} if and only if
$|\inp{\phi_i}{\psi_j}| = 1/\sqrt{d}$, for every $i,j =
1,\ldots,d$. A set $\{\mathcal{B}_1,\ldots,\mathcal{B}_m\}$ of
orthonormal bases in $\Complex^d$ is called a \emph{set of mutually
unbiased bases (MUBs)} if each pair of bases
$\mathcal{B}_i$ and $\mathcal{B}_j$ is mutually unbiased.
\end{definition}
In any dimension $d$, the number of mutually unbiased bases is at
most $d+1$~\cite{boykin:mub}. Explicit constructions are known if
$d$ is a prime power~\cite{boykin:mub,wootters:mub} or a square~\cite{wocjan&beth:mub}. 
We say that a set of unitaries $\{U_s\}$ gives
rise to $|\{U_s\}|$ mutually unbiased bases, 
if those unitaries generate $|\{U_s\}|$ mutually
unbiased bases when applied to the basis vectors of the computational basis.

\subsection{Definitions}\label{definitions}
We now give a more formal description of our problem.
Let $\mY$ and $\mB$ be finite sets and
let $P_{Y,B} = \{p_{yb}\}$ be a probability distribution over 
$\mY \times \mB$. Consider an ensemble 
of quantum states $\ens = \{p_{yb},\rho_{yb}\}$. 
We assume that $\mY$, $\mB$, $\ens$ and $P_{Y,B}$ 
are known to both Alice and Bob. Suppose 
now that Alice chooses $yb \in \mY \times \mB$ 
according to probability distribution $P_{Y,B}$, 
and sends $\rho_{yb}$ to Bob. We can then define the tasks:

\begin{definition}
\emph{State discRimination} \emph{(}$\text{\emph{STAR}}(\ens)$\emph{)}
is the following task for Bob.
Given $\rho_{yb}$, determine $y$. He can perform any
measurement on $\rho_{yb}$ immediately upon receipt.
\end{definition}

\begin{definition}
\emph{State discRimination with Post-measurement Information}
\emph{(}$\text{\emph{PI}}_q\text{\emph{-STAR}}(\ens)$\emph{)} is 
the following task for Bob.
Given $\rho_{yb}$, determine $y$, where Bob can use the following
sources of information in succession.
\begin{enumerate}
\item \label{item1} He can perform any measurement
       on $\rho_{yb}$ immediately upon reception. 
       Afterwards, he can store at most $q$ qubits of quantum
       information about $\rho_{yb}$, and an unlimited amount of
       classical information.
\item \label{item2} After Bob's measurement, Alice announces $b$. 
\item \label{item3} Then, he may measure the remaining $q$ qubits
      depending on $b$ and the measurement outcome obtained in 1.
\end{enumerate}
\end{definition}

We also say that \emph{Bob succeeds at $\text{\emph{STAR}}(\ens)$ or 
$\text{\emph{PI}}_q\text{\text{-STAR}}(\ens)$ with probability
$p$} if and only if $p$ is the average success probability 
$p = \sum_{yb} p_{yb} \Pr\{ \hat{Y} = y | \text{state } \rho_{yb}\}$, 
where $\Pr\{\hat{Y} = y|\text{state }\rho_{yb}\}$ is the 
probability that Bob correctly determines $y$ given $\rho_{yb}$ in the case of STAR, and 
in addition using information sources \ref{item1}, \ref{item2}
and \ref{item3} in the case of PI-STAR.

In this paper, we are interested in the following special case: 
Consider a function $f:\mX \rightarrow \mY$ between finite sets, and
a set of mutually unbiased bases $\mB$ generated 
by a set of unitaries $U_0,U_1,\ldots,U_{|\mB|-1}$
acting on a Hilbert space with basis \{$\ket{x}: x \in \mX$\}.
Take $\ket{\Phi_b^x} = U_b \ket{x}$.
Let $P_X$ and $P_B$ be probability distributions over $\mX$ and $\mB$ respectively.
We assume that $f$, $\mX$, $\mY$, $\mB$, $P_X$, $P_B$, and 
the set of unitaries $\{U_b|b \in \mB\}$
are known to both Alice and Bob. Suppose now that Alice
chooses $x \in \mX$ and  
$b \in \mB$ independently according to probability
distributions $P_X$ and $P_B$ respectively, and sends $\ket{\Phi_b^x}$ to Bob.
Bob's goal is now to compute $y = f(x)$. We thus 
obtain an instance of our problem with states
$\rho_{yb} = \sum_{x \in f^{-1}(y)} P_X(x) \outp{\Phi_b^x}{\Phi_b^x}$. 
We write $\text{STAR}(f)$ and $\text{PI}_q\text{-STAR}(f)$ 
to denote both problems in this special case. We concentrate on the case of 
mutually unbiased bases, as this case is most relevant to our initial goal of analyzing 
protocols for quantum cryptography in the bounded storage model~\cite{serge:bounded}.

Here, we will make use of the basis set
$\mB = \{+,\times,\odot\}$, where $\mB_+ = \{\ket{0},\ket{1}\}$ 
is the computational basis,
$\mB_\times = \{ \frac{1}{\sqrt{2}}(\ket{0} + \ket{1}),
                 \frac{1}{\sqrt{2}}(\ket{0} - \ket{1})\}$
is the Hadamard basis, and
$\mB_\odot = \{ \frac{1}{\sqrt{2}}(\ket{0} + i \ket{1}),
                \frac{1}{\sqrt{2}}(\ket{0} - i\ket{1})\}$ is
what we call the K-basis.
The unitaries that give rise to these bases are $U_+ = \id$, $U_\times = H$ and 
$U_\odot =  K$ with $K = (\id+i \sigma_x)/\sqrt{2}$ respectively.
The Hadamard matrix is given by 
$H = \frac{1}{\sqrt{2}}(\sigma_x + \sigma_z)$.
$\sigma_x$, $\sigma_z$ and $\sigma_y$ are the well-known Pauli matrices.
We generally assume that Bob has no a priori knowledge about the outcome of the function 
and about the value of $b$. This means that
$b$ is chosen uniformly at random from $\mB$, and, in the case of balanced functions, 
that Alice chooses $x$ uniformly at random from $\mX$. More generally,
the distribution is uniform on all $f^{-1}(y)$ and such that each
value $y\in{\cal Y}$ is equally likely.

\subsection{A trivial bound: guessing the basis}

Note that a simple strategy for Bob is to guess the basis, and then measure.
This approach leads to a lower bound on the success probability for 
both $\text{STAR}$ and $\text{PI-STAR}$. In short:
\begin{lemma}\label{guessing}
  Let $P_X(x) = \frac{1}{2^n}$ for all $x \in \01^n$.
  Let $\mB$ denote the set of bases.
  Then for any balanced function $f: \mX \rightarrow \mY$ Bob succeeds at 
  $\text{\emph{STAR}}(f)$ and $\text{\emph{PI}}_0\text{\emph{-STAR}}(f)$ with
  probability at least
  \[
    p_{\text{guess}}=\frac{1}{|\mB|} + \left(1 - \frac{1}{|\mB|}\right) \frac{1}{|\mY|}.
  \]
  \qed
\end{lemma}
\noindent
Our goal is to beat this bound. We show that for $\text{PI-STAR}$, Bob can indeed 
do much better.

\section{No post-measurement information}\label{noinfo}

We first consider the standard case of state discrimination. Here, Alice does not
supply Bob with any additional post-measurement information. Instead, Bob's goal is
to compute $y = f(x)$ immediately. This analysis will enable us to gain interesting insights
into the usefulness of post-measurement information later.

\subsection{Two simple examples}

We now examine two simple one qubit examples of a state discrimination problem,
which we make use of later on. Here, Bob's goal is to learn the value of a bit
which has been encoded in two or three mutually unbiased bases while he does
not know which basis has been used.

\begin{lemma}\label{Breitbart}
Let $x \in \01$, $P_X(x) = \frac{1}{2}$ and $f(x) = x$.
Let $\mB =\{+,\times\}$ with $U_+ = \id$ and $U_\times = H$.
Then Bob succeeds at $\text{\emph{STAR}}(f)$ with probability at most
\benn
p=\frac{1}{2}+\frac{1}{2\sqrt{2}}.
\eenn
There exists a strategy for Bob that achieves $p$.
\end{lemma}
\begin{proof}
The probability of success follows from Theorem~\ref{helstrom}
with $\rho_0 =\frac{1}{2}(\outp{0}{0} +H\outp{0}{0}H)$, 
$\rho_1 =\frac{1}{2}(\outp{1}{1} + H\outp{1}{1}H)$ and $q=1/2$.
\end{proof}

\begin{lemma}\label{Kabuki}
Let $x \in \01$, $P_X(x) = \frac{1}{2}$ and $f(x) = x$.
Let $\mB =\{+,\times,\odot\}$ with $U_+ = \id$, $U_\times = H$ and $U_\odot=K$.
Then Bob succeeds at $\text{\emph{STAR}}(f)$ with probability at most
\benn
p=\frac{1}{2}+\frac{1}{2\sqrt{3}}.
\eenn
There exists a strategy for Bob that achieves $p$.
\end{lemma}
\begin{proof}
The proof is identical to that of Lemma~\ref{Breitbart} using
$\rho_0 =\frac{1}{3}(|0\ran\lan 0| +H|0\ran\lan 0|H+K|0\ran\lan 0|K^\dagger)$,
$\rho_1 =\frac{1}{3}(|1\ran\lan 1| +H|1\ran\lan 1|H+K|1\ran\lan 1|K^\dagger)$,
and $q=1/2$.
\end{proof}

\subsection{An upper bound for all Boolean functions}

We now show that for any Boolean function $f$ and any number of mutually unbiased
bases, the probability that Bob succeeds at $\text{STAR}(f)$ is very limited. 
\begin{theorem}\label{starless0.85}
Let $|\mY|=2$ and let $f$ be a balanced function. 
Then Bob succeeds at $\text{\emph{STAR}}(f)$ with probability at most
\benn
  p = \frac{1}{2}+\frac{1}{2\sqrt{|\mB|}}.
\eenn
In particular, for $|\mB|=2$ we obtain $(1 + 1/\sqrt{2})/2 \approx 0.853$;
for $|\mB|=3$, we obtain $(1 + 1/\sqrt{3})/2 \approx 0.789$.
\end{theorem}
\begin{proof}
The probability of success is given by Theorem~\ref{helstrom}
where for $y \in \01$
\benn
\rho_y =\frac{1}{2^{n-1}|\mB|} \sum_{b=1}^{|\mB|} P_{yb},
\eenn
with $P_{yb} = \sum_{x \in f^{-1}(y)} U_b \outp{x}{x} U_b^{\dagger}$. 
Using the Cauchy-Schwarz inequality we can show that
\benn \label{BoundTrDistWithTrofSquare}
  \|\rho_0-\rho_1\|_1^2 =    [\Tr(|\rho_0-\rho_1| \id)]^2
                        \leq \Tr[(\rho_0-\rho_1)^2] \Tr[\id^2]
                       =     2^n\Tr[(\rho_0-\rho_1)^2],
\eenn
or
\be
  \|\rho_0-\rho_1\|_1 \leq \sqrt{2^n \Tr[(\rho_0-\rho_1)^2]}.
\ee
A simple calculation shows that
\benn
\Tr[(\rho_0-\rho_1)^2]=\frac{4}{2^n|\mB|}.
\eenn
The theorem then follows from the previous equation,
together with Theorem~\ref{helstrom} and
Eq.~(\ref{BoundTrDistWithTrofSquare}).
\end{proof}

\subsection{AND function}

One of the simplest functions to consider is the AND 
function. Recall, that we always assume that 
Bob has no a priori knowledge about the outcome of the 
function. In the case of the AND, this means 
that we are considering a very specific prior: with 
probability $1/2$ Alice will choose the only
string $x$ for which $\text{AND}(x) = 1$. Without any 
post-measurement information, Bob can already
compute the AND quite well.
\begin{theorem}
Let $P_X(x) = \frac{1}{2(2^n-1)}$ for all $x \in \01^n\setminus\{1\ldots1\}$
and $P_X(1\ldots 1) = \frac{1}{2}$.
Let $\mB = \{+,\times\}$ with $U_+ = \id^{\otimes n}$, 
$U_\times = H^{\otimes n}$ and $P_B(+) = P_B(\times) = 1/2$. 
Then Bob succeeds at $\text{\emph{STAR}}(\text{\emph{AND}})$ 
with probability at most
\be\label{OptProbSTARAND}
p=\left \{ \begin{array}{ll}
\frac{1}{2}+\frac{1}{2\sqrt{2}} & \textrm{ if } n=1,\\
1-\frac{1}{2(2^n-1)} & \textrm{ if } n\geq 2.
\end{array}
 \right.
\ee
There exists a strategy for Bob that achieves $p$.
\end{theorem}
\begin{proof}
Let $|c_1\ran=|1\ran^{\otimes n}$ and $|h_1\ran=[H|1\ran]^{\otimes n}$. 
Eq.~(\ref{OptProbSTARAND}) is obtained by substituting
\benn
\rho_0&=\frac{1}{2}\left[\frac{\id-|c_1\ran \lan c_1|}{2^n-1}
                         +\frac{\id-|h_1\ran \lan h_1|}{2^n-1}\right],\\
\rho_1&=\frac{|c_1\ran \lan c_1|+|h_1\ran \lan h_1|}{2},
\eenn
and $q=1/2$ in Theorem \ref{helstrom}.
\end{proof}
In Theorem \ref{PISTARAND}, we will show an optimal bound for 
the case that Bob does indeed
receive the extra information. By comparing the previous
equation with Eq.~(\ref{OptProbPISTARAND}), 
one can see that for $n=1$ announcing the basis does not help. 
However, for $n>1$ we will observe an 
improvement of $[2(2^n+2^{n/2}-2)]^{-1}$.

\subsection{XOR function}

The XOR function provides an example of a Boolean function 
where we observe both the largest advantage 
as well as the smallest advantage in receiving post-measurement 
information: For strings of even length
we will show that without the extra information Bob 
can never do better than guessing the basis. For
strings of odd length, however, he can do quite a bit better. 
Interestingly, it will turn out that in 
this case the post-measurement information is completely useless to him.
We first investigate how well Bob does at $\text{STAR}(\text{XOR})$ for two bases:

\begin{theorem}\label{xorSTAR2MUBS}
Let $P_X(x) = \frac{1}{2^n}$ for all $x \in \01^n$.
Let $\mB = \{+,\times\}$ with $U_+ = \id^{\otimes n}$, 
$U_\times = H^{\otimes n}$ and $P_B(+) = P_B(\times) = 1/2$. 
Then Bob succeeds at $\text{\emph{STAR}}(\text{\emph{XOR}})$ 
with probability at most
\[
  p = \begin{cases} \frac{3}{4} & \text{ if } $n$ \text{ is even}, \\
                    \frac{1}{2}\left(1 + \frac{1}{\sqrt{2}}\right) 
                                & \text{ if } $n$ \text{ is odd}.
      \end{cases}
\]
There exists a strategy for Bob that achieves $p$.
\end{theorem}
\begin{proof}
Our proof works by induction on $n$. 
The case of $n=1$ was addressed in Lemma~\ref{Breitbart}.
Now, consider $n=2$: 
Let
$\sigma^2_0 = \frac{1}{2}(\rho^2_{0+} + \rho^2_{0\times})$ and 
$\sigma^2_1 = \frac{1}{2}(\rho^2_{1+} + \rho^2_{1\times})$, where 
$\rho^2_{0+}$ and $\rho^2_{1+}$ are defined as 
$\rho^n_{o_b b} =
\frac{1}{2^{n-1}} \sum_{x \in \01^n, x \in \text{XOR}^{-1}(o_b)} U_b\outp{x}{x}U_b^{\dagger}$ with
$o_b \in \01$ and $b \in \mB$.
We have $\|\sigma^2_0 - \sigma^2_1\|_1 = 1$.

We now show that the trace distance does not change when we go from 
strings of length $n$ to strings of length $n+2$: 
Define $\rho^n_{yb}$ with $y \in \01, b \in \{+,\times\}$ as in the proof
of Theorem~\ref{xorOddPISTAR2MUBS}, and note that we can express $\rho^{n+2}_{yb}$ as before.
Let $\sigma^n_0 = \frac{1}{2}(\rho^n_{0+} + \rho^n_{0\times})$ and
$\sigma^n_1 = \frac{1}{2}(\rho^n_{1+} + \rho^n_{1\times})$. 
A small calculation shows that 
\begin{equation*}\begin{split}
  \sigma^{n+2}_0 - \sigma^{n+2}_1 &= \frac{1}{8}\bigl[
        (\rho^n_{0+} + \rho^n_{0\times} - \rho^n_{1+} - \rho^n_{1\times})
                                              \otimes \outp{\Phi^+}{\Phi^+} \bigr. \\
     &\phantom{===}
      - (\rho^n_{0+} + \rho^n_{0\times} - \rho^n_{1+} - \rho^n_{1\times}) 
                                              \otimes \outp{\Psi^-}{\Psi^-}        \\
     &\phantom{===}
      + (\rho^n_{0+} + \rho^n_{1\times} - \rho^n_{1+} - \rho^n_{0\times})
                                              \otimes \outp{\Phi^-}{\Phi^-}        \\
     &\phantom{===}
      - \bigl. (\rho^n_{0+} + \rho^n_{1\times} - \rho^n_{1+} - \rho^n_{0\times})
                                              \otimes \outp{\Psi^+}{\Psi^+} \bigr]
\end{split}\end{equation*}
We then get that
\[
  \|\sigma^{n+2}_0 - \sigma^{n+2}_1\|_1
     = \frac{1}{2}(\|\sigma^n_0 - \sigma^n_1\|_1
                   + \|\tilde{\sigma}^n_0 - \tilde{\sigma}^n_1\|_1),
\]
where $\tilde{\sigma}^n_0 = \frac{1}{2}(\rho^n_{0+} + \rho^n_{1\times})$
and $\tilde{\sigma}_n^1 = \frac{1}{2}(\rho^n_{1+} + \rho^n_{0\times})$. 
Consider the unitary $U = \sigma_x^{\otimes n}$ if $n$ is odd, 
and $U = \sigma_x^{\otimes n-1} \otimes \id$
if $n$ is even. It is easy to verify that 
$\sigma^n_0 = U\tilde{\sigma}^n_0U^{\dagger}$ and 
$\sigma^n_1 = U\tilde{\sigma}^n_1U^{\dagger}$. We thus have that 
$\|\sigma^n_0 - \sigma^n_1\|_1 = \|\tilde{\sigma}^n_0 - \tilde{\sigma}^n_1\|_1$ 
and therefore
\[
  \|\sigma^{n+2}_0 - \sigma^{n+2}_1\|_1 = \|\sigma^{n}_0 - \sigma^{n}_1\|_1.
\]

It then follows from Helstrom's theorem~\cite{helstrom} that
the maximum probability to distinguish $\sigma^{n+2}_0$
from $\sigma^{n+2}_1$ and thus compute the XOR of
the $n+2$ bits is given by 
\[
  \frac{1}{2} + \frac{\|\sigma^n_0 - \sigma^n_1\|_1}{4},
\]
which gives the claimed result.
\end{proof}

A similar argument is possible, if we use three mutually unbiased bases.
Intuitively, one might expect Bob's chance of success to drop as we had more bases. 
Interestingly, however, we obtain the same bound of 3/4 if $n$ is even. 

\begin{theorem}\label{xorEvenSTAR3MUBS}
Let $P_X(x) = \frac{1}{2^n}$ for all $x\in\01^n$.
Let $\mB = \{+,\times,\odot\}$ with $U_+ = \id^{\otimes n}$, $U_\times = H^{\otimes n}$,
and $U_\odot=K^{\otimes n}$ with $P_B(+) = P_B(\times) = P_B(\odot) = 1/3$. 
Then Bob succeeds at $\text{\emph{STAR}}(\text{\emph{XOR}})$ with probability at most
\[
  p = \begin{cases} \frac{3}{4} & \text{ if } $n$ \text{ is even}, \\
                    \frac{1}{2}\left(1 + \frac{1}{\sqrt{3}}\right)
                                & \text{ if } $n$ \text{ is odd}.
      \end{cases}
\]
There exists a strategy for Bob that achieves $p$.
\end{theorem}
\begin{proof}
Our proof is very similar to the case of only 2 mutually unbiased bases.
The case of $n=1$ follows from Lemma~\ref{Kabuki}.
This time, we have for $n=2$: 
$\sigma^2_0 = \frac{1}{3}(\rho^2_{0+} + \rho^2_{0\times} + \rho^2_{0\odot})$ and 
$\sigma^2_1 = \frac{1}{3}(\rho^2_{1+} + \rho^2_{1\times} + \rho^2_{1\odot})$.
We have $\|\sigma^2_0 - \sigma^2_1\|_1 = 1$. 

We again show that the trace distance does not change when we go from 
strings of length $n$ to strings of length $n+2$. We can compute
\begin{equation*}\begin{split}
  \sigma^{n+2}_0 - \sigma^{n+2}_1 &= \frac{1}{4}\bigl[
          (\bar{\sigma}^n_1 - \bar{\sigma}^n_0) \otimes \outp{\Phi^+}{\Phi^+} \bigr. \\
    &\phantom{===}-(\hat{\sigma}^n_1 - \hat{\sigma}^n_0) \otimes \outp{\Psi^-}{\Psi^-}\\
    &\phantom{===}+(\tilde{\sigma}^n_1 - \tilde{\sigma}^n_0) \otimes \outp{\Phi^-}{\Phi^-}\\
    &\phantom{===}-\bigl.(\sigma^n_1 - \sigma^n_0) \otimes \outp{\Psi^+}{\Psi^+} \bigr],
\end{split}\end{equation*}
where $\bar{\sigma}^n_1 = (\rho^n_{0+} + \rho^n_{0\times} + \rho^n_{1\odot})/3$,
$\bar{\sigma}^n_0 = (\rho^n_{1+} + \rho^n_{1\times} + \rho^n_{0\odot})/3$,
$\hat{\sigma}^n_1 = (\rho^n_{0+} + \rho^n_{1\times} + \rho^n_{0\odot})/3$,
$\hat{\sigma}^n_0 = (\rho^n_{1+} + \rho^n_{0\times} + \rho^n_{1\odot})/3$,
$\tilde{\sigma}^n_0 = (\rho^n_{1+} + \rho^n_{0\times} + \rho^n_{0\odot})/3$, and
$\tilde{\sigma}^n_0 = (\rho^n_{0+} + \rho^n_{1\times} + \rho^n_{1\odot})/3$.
Consider the unitaries $\bar{U} = \sigma_y^{\otimes n}$,
$\hat{U} = \sigma_x^{\otimes n}$, and $\tilde{U} = \sigma_z^{\otimes n}$ 
if $n$ is odd, and $\bar{U} = \sigma_y^{\otimes n-1} \otimes \id$,
$\hat{U}= \sigma_x^{\otimes n-1} \otimes \id$, 
and $\tilde{U} = \sigma_z^{\otimes n-1} \otimes \id$ 
if $n$ is even. It is easily verified that
$\sigma^n_0 = \bar{U}\bar{\sigma}^n_0\bar{U}^{\dagger}$, 
$\sigma^n_1 = \bar{U}\bar{\sigma}^n_1\bar{U}^{\dagger}$, 
$\sigma^n_0 = \hat{U}\hat{\sigma}^n_0\hat{U}^{\dagger}$, 
$\sigma^n_1 = \hat{U}\hat{\sigma}^n_1\hat{U}^{\dagger}$, 
$\sigma^n_0 = \tilde{U}\tilde{\sigma}^n_0\tilde{U}^{\dagger}$, and
$\sigma^n_1 = \tilde{U}\tilde{\sigma}^n_1\tilde{U}^{\dagger}$. 
We then get that
\[
  \|\sigma^{n+2}_0 - \sigma^{n+2}_1\|_1 = \|\sigma^{n}_0 - \sigma^{n}_1\|_1,
\]
from which the claim follows.
\end{proof}

Surprisingly, if Bob does have some a priori knowledge about the
outcome of the XOR the problem becomes much harder for Bob. 
By expressing the states in the Bell basis and using Helstrom's result,
it is easy to see that if Alice chooses $x \in \01^2$ such that with probability 
$q$, $\text{XOR}(x) = 0$, and with probability $(1-q)$,
$\text{XOR}(x) = 1$, Bob's probability of learning
$\text{XOR}(x)$ correctly is minimized for $q = 1/3$. 
In that case, Bob succeeds with probability
at most $2/3$, which can be achieved  by the trivial 
strategy of ignoring the state he received  and always outputting 1.
This is an explicit example where making a measurement 
does not aid in state discrimination. It has previously
been noted by Hunter~\cite{hunter:nouse}, that such cases 
can exist in mixed state discrimination.

\section{Using post-measurement information}\label{haveinfo}

We are now ready to advance to the core of our problem. Consider an
instance of $\text{PI}_0\text{-STAR}$ with a function
$f:\mX \rightarrow \mY$ and $m=|\mB|$ bases, and some priors $P_X$
and $P_B$ on the sets $\mX$ and $\mB$. Since Bob may not store any
quantum information, all his nontrivial actions are contained in
the first measurement, which must equip him with possible outputs
$o_i \in \mY$ for each basis $i=1,\ldots,m$. In other words, his
most general strategy is a POVM with $|\mY|^m$ outcomes,
each labeled by the strings $o_1,\ldots,o_{m}$ for $o_i \in \mY$ and $m = |\mB|$. 
Once Alice has announced $b$, Bob outputs $\hat{Y} = o_b$.
Here we first prove a general lower bound on the usefulness of
post-measurement information that beats the guessing bound.
Then, we analyze in detail the AND and the XOR function on $n$ bits.

\subsection{A lower bound for balanced functions}\label{lowerBound}

We first give a lower bound on Bob's success probability for any 
balanced function and any number of mutually unbiased bases, by constructing an 
explicit measurement that achieves it.
Without loss of generality, we assume in this section that 
$\mB = [m]$, as otherwise we could consider a lexicographic ordering of $\mB$. 

\begin{theorem}\label{sipiLower}
Let $f:\mX \rightarrow \mY$ be a balanced function, and let $P_X$ and $P_B$ be the uniform
distributions over $\mX$ and $\mB$ respectively. 
Let the set of unitaries $\{U_b|b \in \mB\}$ give rise to 
$|\mB|$ mutually unbiased bases. Choose an encoding such that 
$\forall x,x'\in \mX: \inp{x}{x'} = \delta_{xx'}$.
Then Bob succeeds at $\text{\emph{PI}}_0\text{\emph{-STAR}}(f)$ with probability at least
\[
  p = p_{\text{guess}} + \begin{cases}
                    \frac{|\mY|-1}{|\mY|(|\mY|+3)} & \text{ if } m = 2,             \\
                    \frac{4(|\mY|^2 - 1)}{3\Y(2 + \Y(\Y + 6))} & \text{ if } m = 3, \\
                    -\frac{2}{2\Y}
                    + \frac{2(\Y + m - 1)}{\Y^2 + 3\Y(m-1) + m^2 - 3m + 2}
                                                               & \text{ if } m \geq 4.
                  \end{cases}
\]
where $p_{\text{guess}}$ is the probability that Bob can achieve by guessing the basis as given
in Lemma~\ref{guessing}. In particular, we always have $p > p_{\text{guess}}$.
\end{theorem}
\begin{proof}
Our proof works by constructing a square-root type 
measurement that achieves the lower bound. As explained above,
Bob's strategy for learning $f(x)$ is to perform a 
measurement with $|\mY|^m$ possible outcomes, 
labeled by the strings $o_1,\ldots,o_{m}$ for $o_i \in \mY$ and $m = |\mB|$. 
Once Alice has announced $b$, Bob outputs $f(x) = o_b$.

Take the projector $P_{yb} = \sum_{x \in f^{-1}(y)} \outp{\Phi^x_b}{\Phi^x_b}$
and $\rho_{yb} = \frac{1}{k} P_{yb}$, where $k = |f^{-1}(y)| = |\mX|/|\mY|$. 
Let $M_{o_1,\ldots,o_m}$
denote the measurement operator corresponding to 
outcome $o_1,\ldots,o_m$. Note that
outcome $o_1,\ldots,o_m$ is the correct outcome for 
input state $\rho_{yb}$ if and only if $o_b = y$. 
We can then write Bob's probability of success as
\[
  \frac{1}{m|\mY|} \sum_{o_1,\ldots,o_m \in \mY} \Tr\left(M_{o_1,\ldots,o_m}
                              \left(\sum_{b \in B} \rho_{o_bb}\right)\right).
\]
We will make use of the following measurement:
\[
  M_{o_1,\ldots,o_m} = S^{-\frac{1}{2}}\left(\sum_{b\in \mB}
                                               P_{o_bb}\right)^3 S^{-\frac{1}{2}},
  \text{ with }
  S = \sum_{o_1,\ldots,o_m \in \mY} \left(\sum_{b \in \mB} P_{o_bb}\right)^3.
\]
Clearly, we have $\sum_{o_1,\ldots,o_m \in \mY} M_{o_1,\ldots,o_m} = \id$ and 
$\forall o_1,\ldots,o_{m} \in \mY: M_{o_1,\ldots,o_{m}} \geq 0$ by construction and
thus we indeed have a valid measurement. We first show that $S = c_m\id$:
\begin{equation*}\begin{split}
S &= \sum_{o_1,\ldots,o_m \in \mY} \left(\sum_{b \in \mB} P_{o_bb}\right)^3    \\
  &= \sum_{o_1,\ldots,o_m \in \mY} \sum_{b,b',b'' \in \mB} P_{o_bb}P_{o_{b'}b'}
                                                           P_{o_{b''}b''}      \\
  &= \sum_{o_1,\ldots,o_m \in \mY} \left(\sum_{b} P_{o_bb} 
                          + 2 \sum_{bb', b\neq b'} P_{o_bb}P_{o_{b'}b'}\right. \\
  &\phantom{========}
    \left.+ \sum_{bb', b \neq b'} P_{o_bb}P_{o_{b'}b'}P_{o_bb}
                          + \sum_{bb'b'',b \neq b' \neq b''}
                                     P_{o_bb}P_{o_{b'}b'}P_{o_{b''}b''}\right) \\
  &= \bigl[
      m |\mY|^{m-1} + 2 m (m-1)|\mY|^{m-2} + m (m-1) |\mY|^{m-2}
       + m(m-1)(m-2)|\mY|^{m-3}\bar{\delta}_{2m} \bigr] \id,
\end{split}\end{equation*}
where we have used the assumption that for any $b$, $P_{o_bb}$ is a projector and 
$\sum_{x \in \mX} \outp{\Phi_b^x}{\Phi_b^x} = \id$ which gives 
$\sum_{o_i \in \mY} P_{o_i b_i}
  = \sum_{o_i \in \mY}\sum_{x \in f^{-1}(y)} \outp{\Phi_b^x}{\Phi_b^x} = \id$.
We can then write Bob's probability of success using this particular measurement as
\[
  \frac{1}{c_m k m |\mY|} \sum_{o_1,\ldots,o_{m} \in \mY} 
                          \Tr\left(\left(\sum_{b \in \mB} P_{o_bb}\right)^4\right).
\]
It remains to evaluate this expression. 
Using the circularity of the trace, we obtain
\begin{equation*}\begin{split}
\sum_{o_1,\ldots,o_{m} \in \mY}
   &\Tr\left(\left(\sum_{b \in \mB} P_{o_bb}\right)^4\right)                             \\
   &= \sum_{o_1,\ldots,o_m \in \mY}
        \Tr\left( \sum_{b}P_{o_bb} + 6 \sum_{bb',b\neq b'} P_{o_bb} P_{o_{b'}b'} \right. \\
   &\phantom{==========}
    + 4 \sum_{bb'b'',b\neq b' \neq b''} P_{o_bb}P_{o_{b'}b'}P_{o_{b''}b''}
    + 2 \sum_{bb'b'',b \neq b' \neq b''} P_{o_bb}P_{o_{b'}b'}P_{o_bb}P_{o_{b''}b''}      \\
   &\phantom{==========}
    + \left. \sum_{bb'b''\tilde{b},b\neq b' \neq b'' \neq \tilde{b}}
                             P_{o_bb}P_{o_{b'}b'}P_{o_{b''}b''}P_{o_{\tilde{b}}\tilde{b}}
    + \sum_{bb',b \neq b'} P_{o_bb}P_{o_{b'}b'} P_{o_bb} P_{o_{b'}b'} \right)            \\
   &\geq \bigl[ m |\mY|^{m-1} + 6 m (m-1)|\mY|^{m-2} 
                + 6 m(m-1)(m-2)|\mY|^{m-3} \bar{\delta}_{2m} \bigr.                      \\
   &\phantom{==}
    + \bigl. m (m-1)(m-2)(m-3)|\mY|^{t(m-4)}\bar{\delta}_{2m}\bar{\delta}_{3m} \bigr]\Tr(\id)
             + m(m-1)|\mY|^{m-2}k,
\end{split}\end{equation*}
where we have again used the assumption that for any $b$, $P_{o_bb}$ is a projector and 
$\sum_{x \in \mX} \outp{\Phi_b^x}{\Phi_b^x} = \id$ with $\Tr(\id) = |\mX|$.
For the last term we have used the following: 
Note that $\Tr( P_{o_{b}b}P_{o_{b'}b'}) = k^2/|\mX|$, 
because we assumed mutually unbiased bases. 
Let $r = \rank( P_{o_{b}b}P_{o_{b'}b'})$.
We can then bound $\Tr((P_{o_bb}P_{o_{b'}b'})^2)
 = \sum_i^r \lambda_i( P_{o_{b}b}P_{o_{b'}b'})^2 
 \geq k^4/(|\mX|^2 r) \geq k^3/|\mX|^2 = k/|\mY|^2$, 
where $\lambda_i(A)$ is the $i$-th eigenvalue of a matrix $A$, 
by noting that $r \leq k$ since $\rank(P_{o_bb}) = \rank(P_{o_b'b'}) = k$.
Putting things together we obtain
\[
p \geq \frac{1}{c_m m}\left[G_m(1) 
        + \left(6 + \frac{1}{|\mY|}\right)G_m(2) + 6 G_m(3) + G_m(4)\right],
\]
where $m = |\mB|$, $c_m = G_m(1) + 3G_m(2) + G_m(3)$
and function $G_m: \Natural \rightarrow \Natural$ defined as
$
G_m(i) = \frac{m!}{(m-i)!} |\mY|^{m-i}\prod_{j=2}^{i-1}\bar{\delta}_{mj}.
$ This expression can be simplified to obtain the claimed result.
\end{proof}

Note that we have only used the assumption that Alice uses mutually unbiased bases
in the very last step to say that 
$\Tr(P_{o_bb}P_{o_{b'}b'}) = k^2/|\mX|$. One could generalize 
our argument to other cases by evaluating 
$\Tr(P_{o_bb}P_{o_{b'}b'})$ approximately. 

In the special case $m=|{\cal Y}|=2$ (i.e.~binary function, with
two bases) we obtain:
\begin{corollary}
  \label{eightyfive-percent}
  Let $f:\01^n \rightarrow \01$ be a balanced function and 
  let $P_X(x) = 2^{-n}$ for all $x\in\01^n$.
  Let $\mB = \01$ with $U_0 = \id^{\otimes n}$, 
  $U_1 = H^{\otimes n}$ and $P_B(0)= P_B(1) = 1/2$.
  Then Bob succeeds at $\text{\emph{PI}}_0\text{\emph{-STAR}}(f)$ with probability 
  $p \geq 0.85$.
  \qed
\end{corollary}
Observe that this almost attains the upper bound of $\approx .853$
of Lemma~\ref{Breitbart} in the case of no post-measurement information.
Below (in section~\ref{limitedInfo}) we will show that indeed this bound
can always be achieved when post-measurement information is available.

It is perhaps interesting to note that our general bound 
depends only on the number of function values $|\mY|$
and the number of bases $m$. The number of function inputs 
$|\mX|$ itself does not play a direct role. 

\subsection{Optimal bounds for the AND and XOR function}\label{optimalBound}

We now show that for some specific functions, the 
probability of success can even be much larger.
We hereby concentrate on the case where Alice 
uses two or three mutually unbiased bases 
to encode her input. Our proofs thereby lead to explicit 
measurements. In the following, we again assume that Bob
has no a priori knowledge of the function value. 

\subsubsection{AND function}

\begin{theorem}\label{PISTARAND}
Let $P_X(x) = \frac{1}{2(2^n-1)}$ for all $x \in \01^n\setminus\{1\ldots1\}$
and $P_X(1\ldots 1) = \frac{1}{2}$.
Let $\mB = \{+,\times\}$ with $U_+ = \id^{\otimes n}$, 
$U_\times = H^{\otimes n}$ and $P_B(+) = P_B(\times) = 1/2$. 
Then Bob succeeds at $\text{\emph{PI}}_0\text{\emph{-STAR}}(\text{\emph{AND}})$ 
with probability at most
\be\label{OptProbPISTARAND}
p=\frac{1}{2}\left[2+\frac{1}{2^n+2^{n/2}-2}-\frac{1}{2^n-1}\right].
\ee
There exists a strategy for Bob that achieves $p$.
\end{theorem}
\begin{proof}
To learn the value of $\text{AND}(x)$, Bob uses the same strategy as in the previous section.
He performs a measurement with $4$ possible outcomes, labeled by the strings
$o_+,o_\times$ with $o_+,o_\times \in \01$. 
Once Alice has announced her basis choice $b \in \{+,\times\}$, Bob
outputs $\text{AND}(x) = o_b$. Note that without loss of generality we can assume that Bob's
measurement has only $4$ outcomes, i.e. Bob only stores 2 bits of classical information because
he will only condition his answer on the value of $b$ later on.

Following the approach in the last section, we can write Bob's optimal 
probability of success as a semidefinite program:
\begin{sdp}{maximize}{$\frac{1}{4}\sum_{o_+,o_\times \in \01} 
                       \Tr [b_{o_+ o_\times}M_{o_+o_\times}]$}
&$\forall o_+,o_\times \in \01: M_{o_+o_\times} \geq 0$,\\
&$\sum_{o_+,o_\times \in \01} M_{o_+o_\times} = \id$,
\end{sdp}
where
\begin{eqnarray*}
b_{00}=\rho_{0+}+\rho_{0\times},&
b_{01}=\rho_{0+}+\rho_{1\times},\\
b_{10}=\rho_{1+}+\rho_{0\times},&
b_{11}=\rho_{1+}+\rho_{1\times},
\end{eqnarray*}
with $\forall y \in \01, b \in \{+,\times\}: \rho_{yb} = 
\frac{1}{2} \sum_{x \in \text{AND}^{-1}(y)} U_b\outp{x}{x}U_B^{\dagger}$. 
Consider $\hil_2$, the $2$ dimensional Hilbert space spanned by 
$\ket{c_1}\dfdas\ket{1}^{\otimes n}$ and $\ket{h_1}\dfdas\ket{1_H}^{\otimes n}$.
Let $\ket{c_0}\in \hil_2$ and $\ket{h_0}\in \hil_2$ be the state vectors orthogonal to 
$\ket{c_1}$ and $\ket{h_1}$ respectively. They can be expressed as:
\begin{eqnarray*}
\ket{c_o}&=&\frac{(-1)^{n+1}\ket{c_1}+2^{n/2}\ket{h_1}}{\sqrt{2^n-1}},\\
\ket{h_o}&=&\frac{2^{n/2}\ket{c_1}+(-1)^{n+1}\ket{h_1}}{\sqrt{2^n-1}}.
\end{eqnarray*}
Then $\Pi_\parallel=\outp{c_0}{c_0}+\outp{c_1}{c_1}=\outp{h_0}{h_0}+\outp{h_1}{h_1}$ is 
a projector onto $\hil_2$. Let $\Pi_{\perp}$ be a projector 
onto the orthogonal complement of $\hil_2$.
Note that the $b_{o_+o_\times}$ are all composed of two 
blocks, one supported on $\hil_2$ and the other 
on its orthogonal complement. We can thus write
\be \label{eq:sepbs}
b_{00}&=\frac{2\Pi_\perp}{2^n-1}&+&\frac{\outp{c_0}{c_0}+\outp{h_0}{h_0}}{2^n-1},\\
b_{01}&=\frac{\Pi_\perp}{2^n-1}&+&\left[\frac{\outp{c_0}{c_0}}{2^n-1}+\outp{h_1}{h_1}\right],\\
b_{10}&=\frac{\Pi_\perp}{2^n-1}&+&\left[\frac{\outp{h_0}{h_0}}{2^n-1}+\outp{c_1}{c_1}\right],\\
b_{11}&=0&+&\outp{c_1}{c_1}+\outp{h_1}{h_1}.
\ee
We give an explicit measurement that achieves $p$ and then show
that it is optimal. The full derivation of this measurement can be found in the
appendix. Take
\benn
M_{00}&=\Pi_\perp\\
M_{o_+o_\times}&=\lambda_{o_+o_\times}\outp{\psi_{o_+o_\times}}{\psi_{o_+o_\times}},
\eenn
with 
$\lambda_{01} = \lambda_{10} =(1+\eta)^{-1}$
where 
\benn
\eta &=\left|\frac{1-2 \beta^2+ (-1)^{n+1}2\beta\sqrt{1-\beta^2}\sqrt{2^n-1}}{2^{n/2}}\right|,\\
\ket{\psi_{01}}&=\alpha\ket{c_0}+\beta\ket{c_1},\\
\ket{\psi_{10}}&=\alpha\ket{h_0}+\beta\ket{h_1},
\eenn
with $\alpha$ and $\beta$ real and satisfying $\alpha^2+\beta^2=1$. We also 
set $M_{11} = \id - M_{00} - M_{01} - M_{10}$.
We take 
\benn
\beta = (-1)^n \frac{1}{\sqrt{2^{2n}+2^{\frac{3}{2}n+1}-2^{\frac{n}{2}+1}}}.
\eenn
Putting it all together, we thus calculate Bob's probability of success:
\benn
p=\frac{1}{2}\left[2+\frac{1}{2^n+2^{n/2}-2}-\frac{1}{2^n-1}\right].
\eenn

We now show that this is in fact the optimal measurement for Bob. For this
we will consider the dual of our semidefinite program above:
\begin{sdp}{minimize}{$\Tr(Q)$}
&$\forall o_+,o_\times \in \01: Q\geq \displaystyle \frac{b_{o_+o_\times}}{4}$.
\end{sdp}
Our goal is now to find a $Q$ such that $p = \Tr(Q)$ and $Q$ is dual feasible. 
We can then conclude from the duality of SDP that $p$ is optimal.
Consider
\benn
 Q = &\frac{\Pi_\perp}{2(2^n-1)}
        +\frac{1}{4}\left(\frac{2-{2^{1+n/2}}+{2^{3 n/2}}}{2-3\cdot{2^{n/2}}+{2^{3 n/2}}}\right)
                    (|c_1\ran \lan c_1|+ |h_1\ran \lan h_1|)                                   \\
     &- (-1)^n \frac{1}{4({2^{1-\frac{n}{2}}}+{2^n}-3)}(|c_1\ran \lan h_1|+|c_1\ran \lan h_1|).
\eenn
Now we only need to show that the $Q$ above satisfies 
the constraints, i.e. $\forall o_+,o_\times \in \01:
Q\geq b_{o_+o_\times}/4$. Let $Q_\perp = \Pi_\perp Q \Pi_\perp$ and 
$Q_\parallel = \Pi_\parallel Q \Pi_\parallel$. 
By taking a look at Eq.\ (\ref{eq:sepbs}) one can easily see 
that $Q_\perp \geq \frac{\Pi_\perp b_{o_+o_\times}\Pi_\perp}{4},$ 
so that it is only left to show that 
\benn
  Q_\parallel \geq \frac{\Pi_\parallel b_{o_+o_\times} \Pi_\parallel}{4}, 
                      \textrm{ for }o_+o_\times \in \01, o_+o_\times \neq 00.
\eenn
These are $2 \times 2$ matrices and this can be done straightforwardly.
We thus have $\Tr(Q) = p$ and the result follows from the duality of semidefinite programming.
\end{proof}

It also follows that if Bob just wants to learn the value of a single bit, he can do no better
than what he could achieve without waiting for Alice's announcement of the basis $b$:
\begin{corollary}\label{breitbartPISTAR}
  Let $x \in \01$, $P_X(x) = \frac{1}{2}$ and $f(x) = x$.
  Let $\mB =\{+,\times\}$ with $U_+ = \id$ and $U_\times = H$.
  Then Bob succeeds at $\text{\emph{PI}}_0\text{\emph{-STAR}}(f)$ with probability at most
  \benn
    p = \frac{1}{2}+\frac{1}{2\sqrt{2}}.
  \eenn
  There exists a strategy for Bob that achieves $p$.
  \qed
\end{corollary}

The AND function provides an intuitive example of how Bob can compute the value of a
function perfectly by storing just a single qubit.
Consider the measurement with elements $\{\Pi_\parallel,\Pi_\perp\}$ from the previous section.
It is easy to see that the outcome $\perp$ has zero probability if $\text{AND}(x) = 1$.
Thus, if Bob obtains that outcome he can immediately conclude that $\text{AND}(x) = 0$.
If Bob obtains outcome $\parallel$ then the post-measurement states 
live in a $2$-dimensional Hilbert space ($\hil_2$), and can therefore be stored in a single qubit.
Thus, by keeping the remaining state we can calculate 
the AND perfectly once the basis is announced. 
Our proof in Section~\ref{limitedInfo}, which shows that 
in fact \emph{all} Boolean functions can be
computed perfectly if Bob can store only a single qubit, 
makes use of a very similar effect to
the one we observed here explicitly.

\subsubsection{XOR function}

We now examine the XOR function. This will be useful in order to gain some insight into
the usefulness of post-measurement information later. For strings of even length, there 
exists a simple strategy for Bob even when three mutually unbiased bases are used.

\begin{theorem}\label{xorEvenPISTAR3MUBS}
Let $n \in \Natural$ be even, and
let $P_X(x) = \frac{1}{2^n}$ for all $x \in \01^n$.
Let $\mB =\{+,\times,\odot\}$ with $U_+ = \id^{\otimes n}$, 
$U_\times = H^{\otimes n}$ and $U_\odot=K^{\otimes n}$, where $K=(\id+i\sigma_x)/\sqrt{2}$. 
Then there is a strategy where Bob succeeds at 
$\text{\emph{PI}}_0\text{\emph{-STAR}}(\text{\emph{XOR}})$ with probability 
$p=1$.
\end{theorem}
\begin{proof}
We first construct Bob's measurement for the first 2 qubits, which will allow him
to learn $x_1 \oplus x_2$ with probability 1. Note that the 12 possible states that Alice sends
can be expressed in the Bell basis as follows:
\begin{eqnarray*}
\ket{00} = \frac{1}{\sqrt{2}}(\ket{\Phi^+} + \ket{\Phi^-})
     & H^{\otimes 2}\ket{00} = \frac{1}{\sqrt{2}}(\ket{\Phi^+} + \ket{\Psi^+}) 
          & K^{\otimes 2}\ket{00} = \frac{1}{\sqrt{2}}(\ket{\Phi^-} +i \ket{\Psi^+})\\
\ket{01} = \frac{1}{\sqrt{2}}(\ket{\Psi^+} + \ket{\Psi^-})
     & H^{\otimes 2}\ket{01} = \frac{1}{\sqrt{2}}(\ket{\Phi^-} + \ket{\Psi^-})
          & K^{\otimes 2}\ket{01} = \frac{1}{\sqrt{2}}(i\ket{\Phi^+} + \ket{\Psi^-})\\
\ket{10} = \frac{1}{\sqrt{2}}(\ket{\Psi^+} - \ket{\Psi^-})
     & H^{\otimes 2}\ket{10} = \frac{1}{\sqrt{2}}(\ket{\Phi^-} - \ket{\Psi^-})
          & K^{\otimes 2}\ket{10} = \frac{1}{\sqrt{2}}(i\ket{\Phi^+} - \ket{\Psi^-})\\
\ket{11} = \frac{1}{\sqrt{2}}(\ket{\Phi^+} - \ket{\Phi^-})
     & H^{\otimes 2}\ket{11} = \frac{1}{\sqrt{2}}(\ket{\Phi^+} - \ket{\Psi^+})
          & K^{\otimes 2}\ket{11} = -\frac{1}{\sqrt{2}}(\ket{\Phi^-} - i\ket{\Psi^+}).
\end{eqnarray*}
Bob now simply measures in the Bell basis and records his 
outcome. If Alice now announces that she used the computational
basis, Bob concludes that $x_1 \oplus x_2 = 0$ if the 
outcome is one of $\ket{\Phi^{\pm}}$ and $x_1 \oplus x_2 = 1$ 
otherwise. If Alice announces she used the Hadamard basis, Bob concludes that
$x_1 \oplus x_2 = 0$ if the outcome was one of 
$\{\ket{\Phi^{+}},\ket{\Psi^{+}}\}$ and $x_1 \oplus x_2 = 1$ 
otherwise. Finally, if Alice announces that she used the 
$\odot$ basis, Bob concludes that $x_1 \oplus x_2 = 0$ 
if the outcome was one of $\{\ket{\Phi^{-}},\ket{\Psi^{+}}\}$
and $x_1 \oplus x_2 = 1$ otherwise.
Bob can thus learn the XOR of two bits with probability 1.
To learn the XOR of the entire string, Bob applies this 
strategy to each two bits individually and then computes
the XOR of all answers. 
\end{proof}

Analogously to the proof of Theorem~\ref{xorEvenPISTAR3MUBS}, we obtain:
\begin{corollary}\label{xorEvenPISTAR2MUBS}
  Let $n \in \Natural$ be even, and
  let $P_X(x) = \frac{1}{2^n}$ for all $x \in \01^n$.
  Let $\mB =\{+,\times\}$ with $U_+ = \id^{\otimes n}$ and $U_\times = H^{\otimes n}$.
  Then there is a strategy where Bob succeeds at 
  $\text{\emph{PI}}_0\text{\emph{-STAR}}(\text{\emph{XOR}})$ with probability 
  $p=1$.
  \qed
\end{corollary}

Interestingly, there is no equivalent strategy for 
Bob if $n$ is odd. In fact, as we will show in the next
section, in this case the post-measurement information 
gives no advantage to Bob at all.

\begin{theorem}\label{xorOddPISTAR2MUBS}
Let $n \in \Natural$ be odd, and
let $P_X(x) = \frac{1}{2^n}$ for all $x \in \01^n$.
Let $\mB = \{+,\times\}$ with $U_+ = \id^{\otimes n}$, 
$U_\times = H^{\otimes n}$ and $P_B(+) = P_B(\times) = 1/2$.
Then Bob succeeds at 
$\text{\emph{PI}}_0\text{\emph{-STAR}}(\text{\emph{XOR}})$ 
with probability at most
\[
  p = \frac{1}{2}\left(1 + \frac{1}{\sqrt{2}}\right).
\]
There exists a strategy for Bob that achieves $p$.
\end{theorem}
\begin{proof}
Similar to the proof of the AND function, we can write Bob's optimal
probability of success as the following semidefinite program in terms
of the length of the input string, $n$:
\begin{sdp}{maximize}{$\frac{1}{4}\sum_{o_+,o_\times \in \01}
                       \Tr [b^n_{o_+ o_\times}M_{o_+ o_\times}]$}
&$\forall o_+,o_\times \in \01: M_{o_+ o_\times} \geq 0$,\\
&$\sum_{o_+,o_\times \in \01} M_{o_+ o_\times} = \id$,
\end{sdp}
where
\begin{eqnarray*}
b^n_{o_+ o_\times}&=\rho^n_{o_+ +}+\rho^n_{o_\times \times},
\end{eqnarray*}
and $\rho^n_{o_b b} =
\frac{1}{2^{n-1}} \sum_{x \in \01^n, x \in \text{XOR}^{-1}(o_b)} U_b\outp{x}{x}U_b^{\dagger}$.
The dual can be written as
\begin{sdp}{minimize}{$\frac{1}{4}\Tr(Q^n)$}
&$\forall o_+,o_\times \in \01: Q^n\geq \displaystyle b^n_{o_+o_\times}$.
\end{sdp}
Our proof is now by induction on $n$. For $n=1$, let $Q^1 = 2 p\id$.
It is easy to verify that $\forall o_+,o_\times \in \01:
Q^1 \geq b^1_{o_+o_\times}$ and thus $Q^1$ is a feasible solution of the dual program.

We now show that for $n + 2$, $Q^{n+2} = Q^{n} \otimes \frac{1}{4}\id$ is a feasible
solution to the dual for $n+2$, where $Q^{n}$ is a solution for
the dual for $n$. Note that the
XOR of all bits in the string can be expressed as the 
XOR of the first $n-2$ bits XORed with the XOR of the last two.
We can thus write
\begin{align*}
  \rho_{0+}^{n+2}      &= \frac{1}{2}(\rho^{n}_{0+} \otimes \rho^2_{0+}
                                      + \rho^{n}_{0+} \otimes \rho^2_{1+})           \\
  \rho_{0\times}^{n+2} &= \frac{1}{2}(\rho^{n}_{0\times} \otimes \rho^2_{0\times}
                                      + \rho^{n}_{1\times} \otimes \rho^2_{1\times}) \\
  \rho_{1+}^{n+2}      &= \frac{1}{2}(\rho^{n}_{0+} \otimes \rho^2_{1+}
                                      + \rho^{n}_{1+} \otimes \rho^2_{0+})           \\
  \rho_{1\times}^{n+2} &= \frac{1}{2}(\rho^{n}_{0\times} \otimes \rho^1_{1\times}
                                      + \rho^{n}_{1\times} \otimes \rho^2_{0\times}).
\end{align*}
Now note that we can write
\begin{align*}
  \rho^2_{0+} &= \frac{1}{2}(\outp{00}{00} + \outp{11}{11})
               = \frac{1}{2}(\outp{\Phi^+}{\Phi^+} + \outp{\Phi^-}{\Phi^-}) \\
  \rho^2_{1+} &= \frac{1}{2}(\outp{01}{01} + \outp{10}{10})
               = \frac{1}{2}(\outp{\Psi^+}{\Psi^+} + \outp{\Psi^-}{\Psi^-}).
\end{align*}
It is easy to see that
$\rho^2_{0\times} = H\rho^2_{0+}H
                  = \frac{1}{2}(\outp{\Phi^+}{\Phi^+} + \outp{\Psi^+}{\Psi^+})$
and
$\rho^2_{1\times} = H\rho^2_{1+}H 
                  = \frac{1}{2}(\outp{\Phi^-}{\Phi^-}  + \outp{\Psi^-}{\Psi^-})$.
By substituting from the above equation we then obtain
\begin{equation*}\begin{split}
  b^{n+2}_{00} = \rho^{n+2}_{0+} + \rho^{n+2}_{0\times}
      &= \frac{1}{4}\bigl( (\rho^n_{0+} + \rho^n_{0\times}) \otimes \outp{\Phi^+}{\Phi^+}
                          +(\rho^n_{0+} + \rho^n_{1\times}) \otimes \outp{\Phi^-}{\Phi^-}
                    \bigr. \\
      &\phantom{==}
                    \bigl. (\rho^n_{1+} + \rho^n_{0\times}) \otimes \outp{\Psi^+}{\Psi^+}
                          +(\rho^n_{1+} + \rho^n_{1\times}) \otimes \outp{\Psi^-}{\Psi^-})
                    \bigr) \\
      &\leq \frac{1}{4} Q^n \otimes \id,
\end{split}\end{equation*}
where we have used the fact that $Q^n$ is a feasible solution for the dual for $n$ and
that $\outp{\Phi^+}{\Phi^+} + \outp{\Phi^-}{\Phi^-}
       + \outp{\Psi^+}{\Psi^+} + \outp{\Psi^-}{\Psi^-} = \id$.
The argument for $b^{n+2}_{01}$, $b^{n+2}_{10}$ 
and $b^{n+2}_{11}$ is analogous. Thus $Q^{n+2}$ satisfies all constraints.

Putting things together, we have for odd $n$ that 
$\Tr(Q^{n+2}) = \Tr(Q^n) = \Tr(Q^1)$ and
since the dual is a minimization problem we know 
that $p \leq \frac{1}{4}\Tr(Q^1) = c$ as claimed.
Clearly, there exists a strategy for Bob that 
achieves $p = c$. He can compute the
XOR of the first $n-1$ bits perfectly, as shown 
in Theorem~\ref{xorEvenPISTAR2MUBS}. 
By Corollary~\ref{breitbartPISTAR}
he can learn the value of the remaining $n$-th 
bit with probability $p=c$.
\end{proof}

We obtain a similar bound for three bases:

\begin{theorem}\label{xorOddPISTAR3MUBS}
Let $n \in \Natural$ be odd, and
let $P_X(x) = \frac{1}{2^n}$ for all $x \in \01^n$.
Let $\mB = \{+,\times,\odot\}$ with $U_+ = \id^{\otimes n}$, $U_\times = H^{\otimes n}$ and
$U_\odot=K^{\otimes n}$, where $K=(\id+i\sigma_x)/\sqrt{2}$, with 
$P_B(+) = P_B(\times) = P_B(\odot) = 1/3$. 
Then Bob succeeds at $\text{\emph{PI}}_0\text{\emph{-STAR}}(\text{\emph{XOR}})$
with probability at most
\[
  p = \frac{1}{2}\left(1 + \frac{1}{\sqrt{3}}\right).
\]
There exists a strategy for Bob that achieves $p$.
\end{theorem}
\begin{proof}
The proof follows the same lines as Theorem \ref{xorOddPISTAR2MUBS}. 
Bob's optimal probability of success is:
\begin{sdp}{maximize}{$\displaystyle \frac{1}{6}\sum_{o_+,o_\times,o_\odot \in \01}
                       \Tr [b^n_{o_+ o_\times o_\odot}M_{o_+ o_\times o_\odot}]$}
&$\forall o_+,o_\times,o_\odot \in \01 \in \01: M_{o_+ o_\times o_\odot} \geq 0$,\\
&$\displaystyle\sum_{o_+,o_\times,o_\odot \in \01} M_{o_+ o_\times o_\odot} = \id$,
\end{sdp}
where
\benn
b^n_{o_+ o_\times o_\odot}&=\sum_{b\in \mB}\rho_{o_b b} ,
\eenn
and
\benn
\rho_{o_b b}=\frac{1}{2^{n-1}}\sum_{x\in XOR(o_b)} U_b|x\ran \lan x|U_b^\dagger.
\eenn
The dual can be written as
\begin{sdp}{minimize}{$\frac{1}{6}\Tr(Q^n)$}
&$\forall o_+,o_\times, o_\odot \in \01: Q^n\geq \displaystyle b^n_{o_+ o_\times o_\odot}$.
\end{sdp}
Again, the proof continues by induction on $n$. 
For $n=1$, let $Q^1 = 3 p \id$. It is easy to verify that $\forall o_+,o_\times, o_\odot \in \01:
Q^1 \geq b^1_{o_+ o_\times o_\odot}$ and thus $Q^1$ is a feasible solution of the dual program.
The rest of the proof is done exactly in the same 
way as in Theorem \ref{xorOddPISTAR2MUBS} using  that 
\begin{align*}
  \rho^2_{0\odot} &= \frac{1}{2}(\outp{\Phi^-}{\Phi^-} + \outp{\Psi^+}{\Psi^+}) \\
  \rho^2_{1\odot} &= \frac{1}{2}(\outp{\Psi^-}{\Psi^-} + \outp{\Phi^+}{\Phi^+}).
\end{align*}
\end{proof}

\section[Quantum memory resources for perfect prediction: algebraic framework]%
{Quantum memory resources for perfect prediction:\protect\\ an algebraic framework}
\label{sec:algebra-perfect}

So far, we had assumed that Bob is not allowed to store any qubits and
can only use the additional post-measurement information to improve his guess.
Now, we investigate the case where he has a certain amount of quantum 
memory at his disposal. 
In particular, we present a general
algebraic approach to determine the minimum dimension $2^q$ of quantum
memory needed to succeed with probability $1$ at an instance of
$\text{PI}_q\text{-STAR}(\ens)$ for any ensemble $\ens = \{p_{yb},\rho_{yb}\}$
as long as the individual states for different values of $y$ 
are mutually orthogonal for a fixed $b$, i.e., 
$\forall y\neq z \in \mY\: \Tr(\rho_{yb},\rho_{zb}) =0$.
We are looking for an instrument (w.l.o.g.~maximally
refined) consisting of a family of pure completely positive
maps $\rho \mapsto A \rho A^\dagger$, adding up to a trace preserving
map, such that $\rank \, A \leq 2^q$. This takes care of the memory
bound. The fact that after the announcement of $b$
the remaining state $A \rho_{yb} A^\dagger$ gives full information
about $y$ is expressed by demanding orthogonality of the
different post-measurement states:
\begin{equation}
  \label{pms-orth}
  \forall b \in \mB, \forall y\neq z \in \mY\quad
    A \rho_{yb} A^\dagger A \rho_{zb} A^\dagger = 0.
\end{equation}
Note that here we explicitly allow the possibility that, say, $A \rho_{zb} A^\dagger = 0$:
this means that if Bob obtains outcome $A$ and later learns $b$, 
he can exclude the output value $z$.
What Eq.~(\ref{pms-orth}) also implies is that for all states $\ket{\psi}$
and $\ket{\varphi}$ in the support of $\rho_{yb}$ and $\rho_{zb}$, respectively,
on has $A \ketbra{\psi}{\psi} A^\dagger A \ketbra{\varphi}{\varphi} A^\dagger = 0$,
hence, introducing the support projectors $P_{yb}$ of the $\rho_{yb}$,
we can reformulate Eq.~(\ref{pms-orth}) as
\[
  \forall b \in \mB, \forall y\neq z \in \mY\quad
    A P_{yb} A^\dagger A P_{zb} A^\dagger = 0,
\]
which can equivalently be expressed as
\begin{equation}
  \label{trace-zero}
  \forall b \in \mB, \forall y\neq z \in \mY\quad
    \Tr\bigl( A^\dagger A P_{yb} A^\dagger A P_{zb} \bigr) = 0.
\end{equation}
As expected, we see that only the POVM operators
$M=A^\dagger A$ of the instrument play a role in this condition.
Our conditions can therefore also be written as
$M P_{yb} M P_{zb} = 0$.
From this condition, we now derive the following lemma.
 
\begin{lemma}
  \label{commutation}
  Bob, using an instrument with POVM operators $\{M_i\}$,
  succeeds at \emph{$\text{PI}_q\text{-STAR}$} with probability $1$,
  if and only if
  \begin{enumerate}
    \item for all $i$, $\rank \, M_i \leq 2^q$,
    \item for all $y\in\mY$ and $b \in \mB$, $[M,P_{yb}] = 0$, where
      $P_{yb}$ is the projection on the support of $\rho_{yb}$.
  \end{enumerate}
\end{lemma}
\begin{proof}
  We first show that these two conditions are necessary. Note that
  only the commutation has to be proved:
  let $M$ be a Kraus element from an instrument succeeding
  with probability $1$. Then, for any $y$, $b$, we have by Eq.~(\ref{trace-zero}) that 
  \[
    \Tr\bigl( MP_{yb}M(\id-P_{yb}) \bigr) = 0,
    \text{ hence }
    \Tr\bigl( MP_{yb}MP_{yb} \bigr) = \Tr\bigl( MP_{yb}M \bigr).
  \]
  Thus, by the positivity of the trace on positive operators, 
  the cyclicity of the trace, and $P_{yb}^2 = P_{yb}$ we have that
  \begin{equation*}\begin{split}
    0 &\leq \Tr\bigl( [M,P_{yb}]^\dagger [M,P_{yb}] \bigr)                              \\
      &=    \Tr\bigl( -(MP_{yb}-P_{yb}M)^2 \bigr)                                       \\
      &=    \Tr\bigl( -MP_{yb}MP_{yb} -P_{yb}MP_{yb}M +P_{yb}M^2P_{yb} +MP_{yb}^2M \bigr)
       =    0.
  \end{split}\end{equation*}
  But that means that the commutator $[M,P_{yb}]$ has to be $0$.

  Sufficiency is easy: since the measurement operators commute with
  the states' support projectors $P_{yb}$ (assuming for the moment that
  they are the signals, not the $\rho_{yb}$), and these are orthogonal
  to each other for fixed $b$, the post-measurement states of these
  projectors, $\propto \sqrt{M}P_{yb}\sqrt{M}$ will also be mutually
  orthogonal for fixed $b$. Thus, if Bob learns $b$, he can perform a
  measurement to distinguish the different values of $y$ perfectly.
  The post-measurement states are clearly supported on the support
  of $M$, which can be stored in $q$ qubits. Since Bob's strategy 
  succeeds with probability $1$, it will succeed with probability $1$
  for any states supported in the range of the $P_{yb}$.
\end{proof}

It should be pointed out that the operators $M$ of the instrument
need not commute with the originally given states $\rho_{yb}$.
Nevertheless, the measurement preserves the orthogonality of $\rho_{yb}$ and $\rho_{zb}$ 
with $y\neq z$ for fixed $b$, i.e., $\Tr(\rho_{yb}\rho_{zb}) = 0$.
Now that we know that the POVM operators of the instrument 
have to commute with all the states' support projectors $P_{yb}$, we can invoke
some well-developed algebraic machinery to find the optimal such instrument.

Namely, the $M$ have to come from the commutant
of the operators $P_{yb}$~\cite{bratteli&robinson}. These themselves generate a $*$-subalgebra
${\cal O}$ of the full operator algebra ${\cal B}({\cal H})$ of
the underlying Hilbert space ${\cal H}$, and the structure of
such algebras and their commutants in finite dimension is well understood
(\cite{takesaki:operator-algebras}, Section I.11):
the Hilbert space ${\cal H}$ has a decomposition (i.e., there is an
isomorphism which we write as an equality)
\begin{equation}
  \label{sum-of-products}
  {\cal H} = \bigoplus_j {\cal J}_j \otimes {\cal K}_j
\end{equation}
into a direct sum of tensor products, such that the $*$-algebra ${\cal O}$
and its commutant algebra
${\cal O}' = \bigl\{ M : \forall P\in{\cal O}\ [P,M]=0 \bigr\}$
can be written
\begin{align}
  \label{thisisO}
  {\cal O}  &= \bigoplus_j {\cal B}({\cal J}_j) \otimes \id_{{\cal K}_j},  \\
  \label{thisisOprime}
  {\cal O}' &= \bigoplus_j \id_{{\cal J}_j}     \otimes {\cal B}({\cal K}_j).
\end{align}
 
Koashi and Imoto~\cite{koashi&imoto:operations}, in the context
of finding the quantum operations which leave a set of states
invariant, have described an algorithm to find the commutant ${\cal O}'$,
and more precisely the Hilbert space decomposition~(\ref{sum-of-products}),
of the states $P_{yb}/\Tr P_{yb}$. They show that for this decomposition,
there exist states $\sigma_{j|i}$ on ${\cal J}_j$, 
a conditional probability distribution $\{q_{j|i}\}$, and
states $\omega_j$ on ${\cal K}_j$ which are independent of $i$, such that
we can write them as 
\[
  \forall i\quad \sigma_i = \bigoplus_j q_{j|i}\sigma_{j|i} \otimes \omega_j,
\]
Now, looking at Eq.~(\ref{thisisOprime}), we see that the smallest rank operators
$M \in {\cal O}'$ are of the form $\id_{{\cal J}_j} \otimes \ketbra{\psi}{\psi}$
for some $j$ and $\ket{\psi}\in{\cal K}_j$, and that they are all admissible.
Since we need a family of operators $M$ that are closed to a POVM and thus 
all $j$ have to occur, the minimal quantum memory requirement is
\begin{equation}
  \label{minimal-q}
  \min 2^q = \max_j \dim {\cal J}_j.
\end{equation}
The strategy Bob has to follow is this: For each $j$, pick a basis 
$\{\ket{e_{k|j}}\}$ of the
spaces ${\cal K}_j$ and measure the POVM
$\{\id_{{\cal J}_j}\otimes\ketbra{e_{k|j}}{e_{k|j}}\}$,
corresponding to the decomposition
\[
  {\cal H} = \bigoplus_{jk} {\cal J}_j \otimes \ket{e_{k|j}},
\]
which commutes with the $P_{yb}$. For each outcome, he can store the
post-measurement state in $q$ qubits [as in Eq.~(\ref{minimal-q})], preserving
the orthogonality of the states for different $y$ but fixed $b$. Once he learns
$b$ he can thus obtain $y$ with certainty.

Of course, carrying out the Koashi-Imoto algorithm may not
be a straightforward task in a given situation. Nevertheless, one can
understand the two examples we will present in the following section as special
cases of this general method.

\section{Using post-measurement information and quantum memory}\label{limitedInfo}

We now take a look at two specific cases. First, we show that in fact \emph{all} 
Boolean functions with two bases (mutually unbiased or not) can be computed perfectly 
when Bob is allowed to store just a single qubit. Second, however, we show that there
exist three bases such that for \emph{any balanced} function, Bob must store \emph{all}
qubits to compute the function perfectly. We also give a recipe how to construct such
bases.

\subsection{Using two bases}

For two bases, Bob needs to store only a single qubit to compute any Boolean function
perfectly. As outlined in Section~\ref{sec:algebra-perfect}, 
we need to show that there exists a measurement with the following properties:
First, the posterior states of states corresponding 
to strings $x$ such that $f(x)=0$ are 
orthogonal to the posterior states of states 
corresponding to strings $y$ such that $f(y)=1$. 
Indeed, if this is true and we keep the posterior 
state, then after the basis is announced we 
can distinguish perfectly between both types of states. 
Second, of course, we need that the posterior states are 
supported in subspaces of dimension at most $2$.
The following lemma is the main ingredient in our proof.

\begin{lemma}\label{directsumdecomposition}
Let $f: \{0,1\}^n\to \{0,1\}$ and $P_{0b}=\sum_{x\in f^{-1}(0)}U_b|x\ran \lan x|U_b^\dagger$ 
where $U_0=\id$ and $U_1=U$, then there exists a direct 
sum decomposition of the Hilbert space 
\begin{equation*}
\hil=\bigoplus_{i=1}^m\hil_i, \textrm{ with } \dim \hil_i \leq 2,
\end{equation*}
such that $P_{00}$ and $P_{01}$ can be expressed as
\begin{align*}
  P_{00} &= \sum_{i=1}^m\Pi_i P_{00}\Pi_i, \\
  P_{01} &= \sum_{i=1}^m\Pi_i P_{01}\Pi_i,
\end{align*}
where $\Pi_i$ is the orthogonal projector onto $\hil_i$.
\end{lemma}

\begin{proof}
There exists a basis so that $P_{00}$ and $P_{01}$ can be written as
\benn
P_{00}=\left[\begin{array}{@{}cc@{}}\id_{n_0} & 0_{n_0\times n_1} \\
                            0_{n_1\times n_0} & 0_{n_1\times n_1}\end{array}\right],
P_{01}=\left[\begin{array}{@{}cc@{}}A^{00}_{n_0\times n_0} 
                                              & A^{01}_{n_0\times n_1} \\
            (A^{01})^\dagger_{n_1\times n_0} & A^{11}_{n_1\times n_1}\end{array}\right],
\eenn
where $n_{y}=|f^{-1}(y)|$ is the number of 
strings $x$ such that $f(x)=y$, and we have 
specified the dimensions of the matrix blocks for 
clarity. In what follows these dimensions will be 
omitted. We assume without loss of generality that $n_0 \leq n_1$. 
It is easy to check that, since $P_{01}$ is a projector, it must satisfy
\be \label{P01isaprojector}
A^{00}(\id_{n_0} -A^{00})&=A^{01}{A^{01}}^\dagger,\\
A^{11}(\id_{n_1} -A^{11})&={A^{01}}^\dagger A^{01}.
\ee
Consider a unitary of the following form
\benn
V=\left[\begin{array}{@{}cc@{}}V_0 &0 \\  0 & V_1\end{array}\right],
\eenn
where $V_0$ and $V_1$ are $n_0 \times n_0$ and $n_1 \times n_1$ unitaries respectively. 
Under such a unitary, $P_{00}$ and $P_{01}$ are transformed to:
\benn\label{decomposed}
VP_{00}V^\dagger &= P_{00},\\
VP_{01}V^\dagger &=\left[\begin{array}{@{}cc@{}} 
                            V_0 A^{00}V_{0}^\dagger & V_0 A^{01} V_1^\dagger \\ 
                            (V_0 A^{01} V_1^\dagger)^\dagger & V_1A^{11} V_1^\dagger
                         \end{array}\right].
\eenn
We now choose $V_0$ and $V_1$ from the singular value decomposition
(SVD, \cite[Theorem 7.3.5]{horn&johnson:ma}) of
$A^{01} = V_0^\dagger D V_1$ which gives
\benn
D= V_0 A^{01} V_1^\dagger = \sum_{k=1}^{n_0} d_k |u_k \ran \lan v_k|,
\eenn
where $d_k\geq 0$, $\inp{u_k}{u_l}=\inp{v_k}{v_l}=\delta_{kl}$. 
Since $(A^{01})^\dagger A^{01}$ and 
$A^{01}(A^{01})^\dagger$ are supported in orthogonal 
subspaces, it also holds that $\forall k,l: \inp{u_k}{v_l} = 0$.
Equations (\ref{P01isaprojector}) and (\ref{decomposed}) now give us
\benn
V_0 A^{00}V_0^\dagger(\id_{n_0} -V_0 A^{00}V_0^\dagger)
                &=\sum_{k=1}^{n_0} d_k^2 |u_k \ran \lan u_k|,\\
V_1 A^{11}V_1^\dagger(\id_{n_1} -V_1 A^{11}V_1^\dagger)
                &=\sum_{k=1}^{n_0} d_k^2 |v_k \ran \lan v_k|.
\eenn
Suppose for the time being that all the $d_k$ are 
different. Since they are all non-negative then all 
the $d_k^2$ will also be different and it must hold that
\benn
V_0 A^{00}V_0^\dagger &=\sum_{k=1}^{n_0} a^0_k|u_k \ran \lan u_k|,\\
V_1 A^{11}V_1^\dagger &=\sum_{k=1}^{n_0} a^1_k |v_k \ran \lan v_k|
                         + \sum_{k=n_0+1}^{n_1} a^1_k \outp{\tilde{v}_k}{\tilde{v}_k}
\eenn
for some $a^0_k$, $a^1_k$ and $\ket{\tilde{v}_k}$. 
Note that we can choose $\ket{\tilde{v}_k}$ such that
$\forall k,k',k\neq k': \inp{\tilde{v}_k}{\tilde{v}_{k'}} = 0$ 
and $\forall k,l: \inp{u_k}{\tilde{v}_l} = 0$. 
We can now express $VP_{01}V^\dagger$ as
\benn
VP_{01}V^\dagger = \sum_{k=1}^{n_0}\left[ a^0_k |u_k \ran \lan u_k|
                           + d_k (|u_k \ran \lan v_k| +|v_k \ran \lan u_k|)
                           + a^1_k |v_k \ran \lan v_k|\right]
                   + \sum_{k=n_0+1}^{n_1} a^1_k \outp{\tilde{v}_k}{\tilde{v}_k}.
\eenn
It is now clear that we can choose all
$\hil_k=\spann\{|u_k\ran,|v_k\ran\}$, and $\hil_{k'}=\spann\{\ket{\tilde{v}_{k'}}\}$
which are orthogonal and together add up to $\hil$.

In the case that all the $d_k$ are not different, there 
is some freedom left in choosing $|u_k\ran$ and $|v_k\ran$ 
that still allows us to make $V_0 A^{00}V_0^\dagger$ and 
$V_1 A^{11}V_1^\dagger$ diagonal so that the rest of the 
proof follows in the same way.
\end{proof}

In particular, the previous lemma implies that the posterior 
states corresponding to strings $x$ for which
$f(x) = 0$ are orthogonal to those corresponding to strings $x$ 
for which $f(x) = 1$, which is expressed in the following lemma.
\begin{lemma}\label{posterior-orth}
Suppose  one performs the measurement given by $\{\Pi_i: i\in[m]\}$. 
If the outcome of the measurement is $i$ and the state was $U_b|x\ran$, 
then the posterior state is
\benn
|x,i,b\ran=\frac{\Pi_iU_b|x\ran}{\sqrt{\lan x|U_b^\dagger\Pi_iU_b|x\ran}}.
\eenn
The posterior states satisfy
\begin{eqnarray*}
\forall x\in f^{-1}(0),~x'\in f^{-1}(1),~i\in [m]: \inp{x,i,b}{x',i,b}=0.
\end{eqnarray*}
\end{lemma}
\begin{proof}
The proof follows straightforwardly from that fact that the $\Pi_i$ commute with 
both $P_{00}$ and $P_{01}$ (which follows from Lemma \ref{directsumdecomposition}).
\end{proof}

Now we are ready to prove the main theorem of this section.

\begin{theorem}\label{onequbitisenough}
Let $|\mY|=|\mB|=2$, then there exists a strategy for Bob
such that he succeeds at $\text{\emph{PI}}_1\text{\emph{-STAR}}(\ens)$ with
probability $p = 1$, for any function $f$ and prior $P_X$ on
${\cal X}$ which is uniform on the pre-images $f^{-1}(y)$.
\end{theorem}
\begin{proof}
The strategy that Bob uses is the following:
\begin{itemize}
  \item Bob performs the measurement given by  $\{\Pi_i: i\in[m]\}$.
  \item He will obtain an outcome $i\in [m]$ and 
    store the posterior state which is supported in 
    the at most two-dimensional subspace $\hil_i$.
  \item After the basis $b\in\{0,1\}$ is announced, he
    measures $\{P_{0b}, P_{1b}\}$ and reports the outcome of this measurement.
\end{itemize}
By Lemma~\ref{posterior-orth} this performs with success probability $1$.
\end{proof}

Our result also gives us a better lower bound for all Boolean functions than 
what we had previously obtained in Section~\ref{sipiLower}. Instead of storing
the qubit, Bob now measures it immediately along the 
lines of Lemma~\ref{Breitbart}. It is easy to see that
for one qubit the worst case posterior states to distinguish are in fact 
those in Lemma \ref{Breitbart}.
\begin{corollary} \label{pistarlarger0.85}
  Let $|\mY|=|\mB|=2$, then Bob succeeds at
  $\text{\emph{PI}}_0\text{\emph{-STAR}}(\ens)$ with probability 
  at least $p\geq (1 + 1/\sqrt{2})/2$.
  \qed
\end{corollary}
In particular, our result implies that for the task 
of constructing Rabin-OT in~\cite{serge:bounded}
it is essential for Alice to choose a random function $f$ from
a larger set, which is initially unknown to Bob.

As a final remark, note that in this result, because we succeed with
probability $1$, the prior distributions do not play any role. Likewise, it is
not actually important that the states $\rho_{yb}$ are proportional to
projectors: all that is needed in the most general formulation
of the discrimination problem at the beginning is that for both
$b\in \01$, the states $\rho_{0b}$ and $\rho_{1b}$ are orthogonal.

\subsection{Using three bases}

We have just shown that Bob can compute any Boolean function perfectly when two bases are used. 
However, we now show that 
for any balanced Boolean function there exist three bases, such that Bob needs to store \emph{all} qubits,
in order to compute the function perfectly.
The idea behind our proof is that for a particular choice of three bases, any 
measurement operator that satisfies the conditions set out in Lemma~\ref{commutation} must be proportional 
to the identity. This means that we cannot reduce the number of qubits to be stored by a measurement
and must keep everything. First, we prove the following lemma which we will need in our main proof.

\begin{lemma} \label{misproptoid}
Let $M$ be a selfadjoint matrix which is diagonal in two mutually unbiased bases, then $M$ must be 
proportional to the identity.
\end{lemma}
\begin{proof}
Let $|x\ran$ $|u_x\ran$ $x\in\{1,\ldots,d\}$ be the two MUBs and let $m_x$ be the eigenvalue 
corresponding to $|x\ran$ and $|u_x\ran$, then we can write
\benn
 M=\sum_{x=1}^d m_x |x\ran \lan x|=\sum_{x^\pr=1}^d m_{x^\pr} |u_{x^\pr}\ran \lan u_{x^\pr}|.
\eenn
From the previous equation, it follows that
\benn
 \lan x|M|x \ran = m_x=\sum_{{x^\pr}=1}^d m_{x^\pr} |\lan u_{x^\pr}|x\ran|^2=\frac{1}{d} \Tr M,
\eenn
which implies the desired result.
\end{proof}

We are now ready to prove the main result of this section.

\begin{theorem}
Let $|\mY|=2$ and $|\mB|=3$, then for any balanced function $f$ and prior $P_X$ on
${\cal X}$ which is uniform on the pre-images $f^{-1}(y)$, there exist three bases 
such that Bob succeeds at $\text{\emph{PI}}_q\text{\emph{-STAR}}(\ens)$ with
probability $p = 1$ if and only if $q=\log d$.
\end{theorem}
\begin{proof}
Let $P_{00}=\sum_{x\in f^{-1}(0)}|x\ran \lan x|$, $P_{01}=U_1 P_{00}U_1^\dagger$ and $
P_{02}=U_2 P_{00}U_2^\dagger$. Also, let $s:f^{-1}(0)\to f^{-1}(1)$ be a bijective map, and let $s_x=s(x)$. 
By a reordering of the basis, $P_{00}$, $U_1$ and $U_2$ can be written as
\benn
P_{00}=\left[\begin{array}{@{}cc@{}}\id & 0 \\
                            0 & 0\end{array}\right],
U_1=\left[\begin{array}{@{}cc@{}}U_1^{00} 
                                              & U_1^{01} \\
            U_1^{10} & U_1^{11}\end{array}\right],
U_2=\left[\begin{array}{@{}cc@{}}U_2^{00} 
                                              & U_2^{01} \\
            U_2^{10} & U_2^{11}\end{array}\right],
\eenn
where all the blocks are of size $(d/2) \times (d/2)$. $P_{01}$ and $P_{02}$  then take the following form:
\benn
P_{01}=\left[\begin{array}{@{}cc@{}}U_1^{00}{U_1^{00}}^\dagger
                                              & U_1^{00}{U_1^{10}}^\dagger \\
            (U_1^{00}{U_1^{10}}^\dagger)^\dagger & U_1^{10}{U_1^{10}}^\dagger\end{array}\right],
P_{02}=\left[\begin{array}{@{}cc@{}}U_2^{00}{U_2^{00}}^\dagger
                                              & U_2^{00}{U_2^{10}}^\dagger \\
            (U_2^{00}{U_2^{10}}^\dagger)^\dagger & U_2^{10}{U_2^{10}}^\dagger\end{array}\right].
\eenn

It follows from Lemma~\ref{commutation}, that we only need to prove that 
$[M,P_{00}]=[M,P_{01}]=[M,P_{02}]=0$ implies that $M$ must be proportional to the identity. Write
\benn
M=\left[\begin{array}{@{}cc@{}}M^{00} 
                                              & M^{01} \\
            (M^{01})^\dagger & M^{11}\end{array}\right].
\eenn
Commutation with $P_{00}$ implies $M^{01}=0$. Commutation with $P_{01}$ and $P_{02}$ implies
\begin{eqnarray}
\label{condsforM1}
{[}M^{00},U_1^{00}{U_1^{00}}^\dagger{]}&=&{[}M^{00},U_2^{00}{U_2^{00}}^\dagger{]}=0,\\
\label{condsforM2}
[M^{11},U_1^{10}{U_1^{10}}^\dagger]&=&[M^{11},U_2^{10}{U_2^{10}}^\dagger]=0,\\
\label{last1Cond}
M^{00}(U_1^{00}{U_1^{10}}^\dagger)&=&(U_1^{00}{U_1^{10}}^\dagger)M^{11},\\
\label{last2Cond}
M^{00}(U_2^{00}{U_2^{10}}^\dagger)&=&(U_2^{00}{U_2^{10}}^\dagger)M^{11}.
\end{eqnarray}
We choose $U_1$ and $U_2$ in the following way:
\benn
U_1 &=\sum_{x\in f^{-1}(0)}\left[a_x (|x\ran \lan x|+|s_x\ran \lan s_x|) +\sqrt{1-a_x^2}(|x\ran \lan s_x|-|s_x\ran \lan x|)\right],\\
U_2 &=\sum_{x\in f^{-1}(0)}\left[a_x (|u_x\ran \lan u_x|+|v_x\ran \lan v_x|) +\sqrt{1-a_x^2}(|u_x\ran \lan v_x|-|v_x\ran \lan u_x|)\right],
\eenn
with $a_x\in [0,1]$, satisfying $a_x=a_{x^\pr}$ if and only if $x=x^\pr$. 
Furthermore, choose $\ket{u_x}$ and $\ket{v_x}$ such that 
\benn
\forall x,x^\pr \in f^{-1}(0),~ \lan x|v_{x^\pr}\ran=\lan s_x|u_{x^\pr}\ran=0,~
|\lan x|u_{x^\pr}\ran|^2=|\lan s_x|v_{x^\pr}\ran|^2=(d/2)^{-1}.
\eenn

With this choice for $U_1$ and $U_2$ we have that
\benn
U_1^{00}{U_1^{00}}^\dagger&=\sum_{x\in f^{-1}(0)} a_x^2 \outp{x}{x},\\
U_2^{00}{U_2^{00}}^\dagger&=\sum_{x\in f^{-1}(0)} a_x^2 \outp{u_x}{u_x},
\eenn
i.e., $\{\ket{x}\}$ and $\{\ket{u_x}\}$ form an eigenbasis for $U_1^{00}{U_1^{00}}^\dagger$ and 
$U_2^{00}{U_2^{00}}^\dagger$ respectively. Furthermore, since all the $a_x^2$ are different, 
the eigenbases are unique. Now, using Eq.\ (\ref{condsforM1}), we see that $M^{00}$ must commute 
with both $U_1^{00}{U_1^{00}}^\dagger$  and $U_2^{00}{U_2^{00}}^\dagger$, and since their eigenbases 
are unique, it must be true that $M^{00}$ is diagonal in both  $\{\ket{x}\}$ and $\{\ket{u_x}\}$. Using the 
result of Lemma \ref{misproptoid} it follows that $M^{00}=m_0 \id_{d/2}$. In exactly the same way 
we can prove that $M^{11}=m_1 \id_{d/2}$ using Eq.\ (\ref{condsforM2}). 
It remains to prove that $m_0=m_1$, which follows directly from either Eq.\ (\ref{last1Cond}) or Eq.\ (\ref{last2Cond}).
\end{proof}

From our proof it is clear how to construct appropriate $U_1$ and $U_2$. Note, however, that whereas
we know that for such unitaries Bob must store all qubits in order to compute the value of the function
perfectly, it remains unclear how close he can come to computing the function perfectly. In particular,
he can always choose two of the three bases, and employ the strategy outlined in the previous section:
he stores the one qubit that allows him to succeed with probability $1$. If he gets the third basis 
then he just flips a coin. In this case, he is correct with probability $2/3 + 1/(3 \cdot 2) = 5/6$ 
for a balanced function and a uniform prior.

\section{Conclusion and open questions}
\label{conclusion}
We have introduced a new state discrimination problem, motivated by
cryptography: discrimination with extra information about the state
after the measurement, or, more generally, after a quantum memory bound
applies.  We have left most general questions open, but we found fairly complete
results in the case of guessing $y = f(x)$ with mutually unbiased encodings.

We have shown that storing just a single qubit allows Bob to succeed at
PI-STAR perfectly for \emph{any} Boolean function and any two bases.
On the contrary, we showed how to construct \emph{three} bases such that
Bob needs to store \emph{all} qubits in order to compute the function
perfectly.

We have also given an explicit strategy for two functions, namely the AND 
and the XOR. More generally, it would be interesting to find out,
how many qubits Bob needs to store to compute $f(x)$ 
perfectly for any function $f: \mX \rightarrow \mY$
in terms of the number of outputs $|\mY|$ and the number of bases $|\mB|$.
It should be clear that the algebraic techniques of
Section~\ref{sec:algebra-perfect} allow us to answer
these questions for any given function in principle. However, so far, we 
have not been able to obtain general structures for wider
classes of functions.

Our results imply that in existing protocols in the bounded quantum storage 
model~\cite{serge:bounded} we cannot restrict ourselves to a single fixed function $f$.
However, a great challenge arises in considering more than one function, where
$f$ is also announced after the memory bound applies~\cite{serge:bounded}. 

In general, it is an interesting problem to consider 
when post-measurement information is useful 
and how large the advantage can be for Bob. In the 
important case of two mutually unbiased bases and 
balanced functions, we have shown (Theorem \ref{starless0.85} 
and Corollary \ref{pistarlarger0.85}) that there exists a clear separation 
between the case where Bob gets the post-measurement information (PI-STAR) 
and when he does not (STAR).  Namely, for any such function, Bob's
optimal success probability is never larger
than $(1 + 1/\sqrt{2})/2 \approx 0.853$ for STAR and always 
at least as large as the same number for PI-STAR.

In some cases the gap between STAR and PI-STAR can be 
more dramatic. The XOR function on strings of even
length with two mutually unbiased bases is one of 
these cases. We have shown that in this case the advantage can be 
maximal. Namely, \emph{without} the extra information 
Bob can never do better than guessing the 
basis, \emph{with} it however, he can compute the 
value of the function perfectly. This contrasts with the 
XOR function on strings of odd length, where the optimal success probabilities 
of STAR and PI-STAR are both $(1 + 1/\sqrt{2})/2$ 
and the post-measurement information is completely useless for Bob. 
It would be interesting to see, how large the
gap between STAR and PI-STAR can be for any function 
$f: \01^n \rightarrow \01^k$ where $k > 2$.
It would also be nice to show a general lower bound for non-balanced
functions or a non-uniform prior. As the example for $3$ bases
showed, the uniform prior is not necessarily the one that leads to the largest gap, and
thus the prior can play an important role.
Another generalization would be to consider functions 
of the form $f: [d]^n \rightarrow [d]^k$. 

We close by pointing out a potentially interesting connection
to the problem of information locking with mutually unbiased
bases~\cite{terhal:locking} and random bases~\cite{winter:randomizing}.
There, the objective is not so much to
obtain an accurate guess of the value $y = f(x)$, as to maximize
the (classical) mutual information at the end. In locking, we distinguish
measurement with basis information, analogous to our $\text{PI}_q\text{-STAR}$
with $q=n$, and without (or rather only after the measurement),
corresponding to $\text{PI}_0\text{-STAR}$.
From a classical perspective, it is surprising that
the difference in attainable accessible information between
$q=n$ and $q=0$ can be much larger than the information contained in 
a message specifying which basis was used.
In our scenario, we are not interested in locking a string 
$x$, but in locking $f(x)$ for a fixed function $f$.
The strength of the observed locking effect depends on the ratio of the number 
of values $f$ can take and the number of bases used. The dependence on the number of bases
carries over to information locking~\cite{winter:randomizing}, but see
the cautionary tale of~\cite{ballester&wehner:locking}.
It would be interesting to generalize information locking to intermediate
values of $q$, but it seems that we first need to understand the
intricate conditions the \emph{bases} have to meet to ensure locking
in the first place.

\section*{Acknowledgments}
We thank Harry Buhrman for his persistent interest in 
the present investigation, various discussions, and 
his suggestion that our problem also has applications to communication
complexity. 
Thanks also to Ronald de Wolf for helpful comments on an earlier
version of the manuscript.
MB and SW are supported by an NWO vici grant 2004-2009
and by the EU project QAP (IST-2005-15848).
AW is supported by the U.K.~EPSRC's ``QIP IRC'', the EU project QAP,
and by a University of Bristol Research Fellowship.

\appendix

\section*{A. Optimal measurement for the AND function}
\label{app:AND}
For the interested reader, we present the line of thought of deriving the
optimal measurement of computing the AND function, when we are allowed to use post-measurement
information. Here, Bob is not allowed to store any qubits.

Supported by numerical calculations we construct the following measurement:
\benn
M_{00}&=\Pi_\perp\\
M_{o_+o_\times}&=\lambda_{o_+o_\times}\outp{\psi_{o_+o_\times}}{\psi_{o_+o_\times}},
\eenn
for some $\ket{\psi_{o_+o_\times}}\in\hil_2,~\textrm{for }o_+o_\times \neq 00$ chosen later.
Since $\Pi_\parallel\ket{\psi_{o_+o_\times}}=\ket{\psi_{o_+o_\times}}$ 
we can express Bob's probability of success using such a measurement as
\be\label{probSucc}
p = \frac{1}{4}\left[\Tr[b_{00}\Pi_\perp]+\sum_{o_+o_\times \in \01, o_+o_\times\neq 00} 
\Tr [\Pi_\parallel b_{o_+o_\times}\Pi_\parallel M_{o_+o_\times}]\right].
\ee
When we project $b_{01}$, $b_{10}$ and $b_{11}$ onto $\hil_2$ we obtain
\be \label{eq:projections}
\begin{array}{lll}
\ds \Pi_\parallel\rho_{0+}\Pi_\parallel=\frac{\outp{c_0}{c_0}}{2^n-1},&\qquad&
\ds \Pi_\parallel\rho_{0\times}\Pi_\parallel=\frac{\outp{h_0}{h_0}}{2^n-1},\\
\\
\Pi_\parallel\rho_{1+}\Pi_\parallel=\rho_{1+}=\outp{c_1}{c_1},&\qquad&  
\Pi_\parallel\rho_{1\times}\Pi_\parallel=\rho_{1\times}=\outp{h_1}{h_1}.
\end{array}
\ee
Substituting into Equation~\ref{probSucc} we then get
\benn
p=\frac{1}{4}&\left[ 2 \frac{2^n-2}{2^n-1}
                     +\frac{\bra{c_0}M_{01}\ket{c_0}}{2^n-1}+\bra{h_1}M_{01}\ket{h_1}
               \right. \\
             &\left. +\frac{\bra{h_0}M_{10}\ket{h_0}}{2^n-1}+\bra{c_1}M_{10}\ket{c_1}
                     +\bra{c_1}M_{11}\ket{c_1} + \bra{h_1}M_{11}\ket{h_1}\right].
\eenn
Now using that $M_{11}=\Pi_\parallel-M_{01}-M_{10}$ we get
\benn
p=\frac{1}{4}\left[ 2 \frac{2^n-2}{2^n-1}+\frac{\bra{c_0}M_{01}\ket{c_0}}{2^n-1}
                    -\bra{c_1}M_{01}\ket{c_1}
                    +\frac{\bra{h_0}M_{10}\ket{h_0}}{2^n-1}-\bra{h_1}M_{10}\ket{h_1}+2\right].
\eenn

We now show how to choose $\ket{\psi_{o_+o_\times}}$. We take 
$\lambda = \lambda_{01}=\lambda_{10}$. Then $s_1$ and $s_2$, 
the only possible nonzero eigenvalues of 
\benn
M_{01} + M_{10} = \lambda (|\psi_{01}\ran\lan\psi_{01}|+|\psi_{10}\ran\lan\psi_{10}|),
\eenn
satisfy
\benn
s_1+s_2&=2 \lambda\\
s_1^2+s_2^2&=2 \lambda^2(1+|\inp{\psi_{01}}{\psi_{10}}|^2).
\eenn
From these two equations one gets 
\benn
s_1&=\lambda(1+\eta),\\
s_2&=\lambda(1-\eta),
\eenn
where $\eta~\e^{\iu \phi}=\inp{\psi_{10}}{\psi_{01}}$%
\footnote{In the present context $\e^{\iu \phi}=\pm1$}. 
Recall that we need to have $M_{01} + M_{10} + M_{11} = \Pi_\parallel$,
which is equal to the identity on $\hil_2$.
We therefore want one of the eigenvalues of $M_{01} + M_{10}$ 
to be 1 and the other one smaller than 1. 
So we must choose $\lambda=(1+\eta)^{-1}$. We then also need
\benn
\lambda_{11}&=\frac{2 \eta}{1+\eta},\\
\ket{\psi_{11}}&=\frac{\ket{\psi_{01}}-\e^{\iu \phi}\ket{\psi_{10}}}{\sqrt{2(1-\eta)}}.
\eenn
We then take, supported by the symmetry of the problem,
\benn
\ket{\psi_{01}}&=\alpha\ket{c_0}+\beta\ket{c_1},\\
\ket{\psi_{10}}&=\alpha\ket{h_0}+\beta\ket{h_1},
\eenn
with $\alpha$ and $\beta$ real and satisfying 
$\alpha^2+\beta^2=1$. We have now that
\benn
\inp{\psi_{10}}{\psi_{01}}=(-1)^{n+1}\frac{\alpha^2-\beta^2}{2^{n/2}}
                            + 2\alpha \beta \sqrt{1-\frac{1}{2^n}}.
\eenn
Now $p$ becomes
\benn
p&=\frac{1}{2}\left[ \frac{2^n-2}{2^n-1}+\frac{1}{1+\eta}\left(\frac{\alpha^2}{2^n-1}
                     - \beta^2\right)+1\right],\\
 &=\frac{1}{2}\left[ \frac{2^n-2}{2^n-1}+\frac{1}{1+\eta}\frac{1-2^n\beta^2 }{2^n-1}+1\right],
\eenn
and
\benn
\eta&=\left|\frac{\alpha^2-\beta^2+ (-1)^{n+1}2\alpha \beta\sqrt{2^n-1}}{2^{n/2}}\right|\\
    &=\left|\frac{1-2 \beta^2+ (-1)^{n+1}2\beta\sqrt{1-\beta^2}\sqrt{2^n-1}}{2^{n/2}}\right|,
\eenn
where without loss of generality we have chosen $\alpha$ to be positive. 
We want $\eta$ to be small. 
It is easy to see that we would like to take $\beta = (-1)^n |\beta'|$, 
for some real $\beta'$.  A simple calculation shows that then to minimize $\eta$ we should
choose
\benn
|\beta'|=\frac{1}{\sqrt{2^{2n}+2^{\frac{3}{2}n+1}-2^{\frac{n}{2}+1}}}.
\eenn


\small
\begin{thebibliography}{10}

\bibitem{aharonov&englert:meanking}
Y.~Aharonov and B.-G. Englert.
\newblock The mean king's problem: Prime degrees of freedom.
\newblock {\em Physics Letters A}, 284:1--5, 2001.
\newblock quant-ph/0101134.

\bibitem{ballester&wehner:locking}
M.~Ballester and S.~Wehner.
\newblock Locking with three mutually unbiased bases.
\newblock quant-ph/0606244, 2006.

\bibitem{ban:symmetric}
M.~Ban, K.~Kurokawa, R.~Momose, and O.~Hirota.
\newblock Optimum measurements for discrimination among symmetric quantum
  states and parameter estimation.
\newblock {\em International Journal of Theoretical Physics}, 36(6):1269--1288,
  1997.

\bibitem{boykin:mub}
S.~Bandyopadhyay, P.~O. Boykin, V.~P. Roychowdhury, and F.~Vatan.
\newblock A new proof for the existence of mutually unbiased bases.
\newblock {\em Algorithmica}, 34(4):512--528, 2002.

\bibitem{barnett:symmetric}
S.~M. Barnett.
\newblock Minimum-error discrimination between multiply symmetric states.
\newblock {\em Physical Review A}, 64:030303, 2001.

\bibitem{bergou:filtering0}
J.~Bergou, U.~Herzog, and M.~Hillery.
\newblock Quantum state filtering and discrimination between sets of boolean
  functions.
\newblock {\em Physical Review Letters}, 90:257901, 2003.

\bibitem{bergou:survey}
J.~Bergou, U.~Herzog, and M.~Hillery.
\newblock Discrimination of quantum states.
\newblock In M.~Paris and J.~Rehacek, editors, {\em Quantum State Estimation},
  volume~3, pages 417--465. Springer, Berlin, 2004.

\bibitem{bergou:filtering}
J.~Bergou, U.~Herzog, and M.~Hillery.
\newblock Optimal unambiguous filtering of a quantum state: An instance in
  mixed state discrimination.
\newblock {\em Physical Review A}, 71:042314, 2005.

\bibitem{bergou:filtering2}
J.~Bergou and M.~Hillery.
\newblock Quantum-state filtering applied to the discrimination of boolean
  functions.
\newblock {\em Physical Review A}, 72:012302, 2005.

\bibitem{boyd:book}
S.~Boyd and L.~Vandenberghe.
\newblock {\em Convex Optimization}.
\newblock Cambridge University Press, 2004.

\bibitem{bratteli&robinson}
O.~Bratteli and D.~W. Robinson.
\newblock {\em Operator algebras and quantum statistical mechanics. 1. C${}^*$-
  and W${}^*$-algebras, symmetry groups, decomposition of states}.
\newblock Texts and Monographs in Physics. Springer Verlag, 2nd edition, 1987.

\bibitem{serge:bounded}
I.~Damgaard, S.~Fehr, L.~Salvail, and C.~Schaffner.
\newblock Cryptography in the {B}ounded {Q}uantum-{S}torage {M}odel.
\newblock In {\em Proceedings of 46th IEEE FOCS}, pages 449--458, 2005.
\newblock quant-ph/0508222v2.

\bibitem{kretschmann&werner:impossible}
G.~M. D'Ariano, D.~Kretschmann, D.~Schlingemann, and R.~F. Werner.
\newblock Quantum bit commitment: the possible and the impossible.
\newblock quant-ph/0605224, 2006.

\bibitem{terhal:locking}
D.~P. DiVincenzo, M.~Horodecki, D.~W. Leung, J.~A. Smolin, and B.~W. Terhal.
\newblock Locking classical correlation in quantum states.
\newblock {\em Physical Review Letters}, 92:067902, 2004.
\newblock quant-ph/0303088.

\bibitem{eldar:sdp}
Y.~Eldar.
\newblock A semidefinite programming approach to optimal unambiguous
  discrimination of quantum states.
\newblock {\em {IEEE} Transactions on Information Theory}, 49:446--456, 2003.

\bibitem{eldar:pgm}
Y.~Eldar and G.~Forney.
\newblock On quantum detection and the square-root measurement.
\newblock {\em {IEEE} Transactions on Information Theory}, 47:858--872, 2001.

\bibitem{eldar:sdpDetector}
Y.~Eldar, A.~Megretski, and G.~Verghese.
\newblock Designing optimal quantum detectors via semidefinite programming.
\newblock {\em {IEEE} Transactions on Information Theory}, 49:1007--1012, 2003.

\bibitem{eldar:symmetric}
Y.~Eldar, A.~Megretski, and G.~Verghese.
\newblock Optimal detection of symmetric mixed quantum states.
\newblock {\em {IEEE} Transactions on Information Theory}, 50:1198--1207, 2004.

\bibitem{hausladen:pgm}
P.~Hausladen and W.~K. Wootters.
\newblock A pretty good measurement for distinguishing quantum states.
\newblock {\em Journal of Modern Optics}, 41:2385--2390, 1994.

\bibitem{winter:randomizing}
P.~Hayden, D.~Leung, P.~Shor, and A.~Winter.
\newblock Randomizing quantum states: Constructions and applications.
\newblock {\em Communications in Mathematical Physics}, 250(2):371--391, 2004.
\newblock quant-ph/0307104.

\bibitem{helstrom}
C.~W. Helstrom.
\newblock Quantum detection and estimation theory.
\newblock {\em J. Stat. Phys.}, 1(2):231--252, 1969.

\bibitem{holevo:maxState}
A.~S. Holevo.
\newblock Statistical decision theory for quantum systems.
\newblock {\em Journal of Multivariate Analysis}, 3(337), 1973.

\bibitem{horn&johnson:ma}
R.~A. Horn and C.~R. Johnson.
\newblock {\em Matrix Analysis}.
\newblock Cambridge University Press, 1985.

\bibitem{hunter:nouse}
K.~Hunter.
\newblock Measurement does not always aid state discrimination.
\newblock {\em Physical Review A}, 68:012306, 2003.

\bibitem{klappenecker&roetteler:meanking}
A.~Klappenecker and M.~R\"otteler.
\newblock {\frakfamily New Tales of the Mean King}.
\newblock quant-ph/0502138, 2005.

\bibitem{koashi&imoto:operations}
M.~Koashi and N.~Imoto.
\newblock Operations that do not disturb partially known quantum states.
\newblock {\em Physical Review A}, 66:022318, 2002.
\newblock quant-ph/0101144.

\bibitem{lo&chau:bitcom}
H-K. Lo and H.~F. Chau.
\newblock Is quantum bit commitment really possible?
\newblock {\em Physical Review Letters}, 78:3410--3413, 1997.
\newblock quant-ph/9603004.

\bibitem{mayers:trouble}
D.~Mayers.
\newblock The trouble with quantum bit commitment.
\newblock quant-ph/9603015, 1996.

\bibitem{mochon:pgm}
C.~Mochon.
\newblock A family of generalized `pretty good' measurements and the
  minimal-error pure-state discrimination problems for which they are optimal.
\newblock quant-ph/0506061.

\bibitem{takesaki:operator-algebras}
M.~Takesaki.
\newblock {\em Theory of Operator Algebras. I}.
\newblock Springer Verlag, 1979.

\bibitem{wang:classification}
M.~Wang and F.~Yan.
\newblock Conclusive quantum state classification.
\newblock quant-ph/0605127.

\bibitem{wocjan&beth:mub}
P.~Wocjan and T.~Beth.
\newblock New Construction of Mutually Unbiased Bases in Square Dimensions.
\newblock {\em QIC}, 5(2):129--158,2005.

\bibitem{wootters:mub}
W.~K. Wootters and B.~Fields.
\newblock Optimal state-determination by mutually unbiased measurements.
\newblock {\em Ann. Phys.}, 191(368), 1989.

\bibitem{yuen:maxState}
H.~P. Yuen, R.~S. Kennedy, and M.~Lax.
\newblock Optimum testing of multiple hypotheses in quantum detection theory.
\newblock {\em {IEEE} Transactions on Information Theory}, 21, 1975.

\end{thebibliography}
\end{document}